\documentclass[11pt,a4paper]{article}
\pdfoutput=1
\usepackage{jheppub}
\usepackage[utf8]{inputenc}

\usepackage{color}
\usepackage{soul}
\usepackage{amsmath}
\usepackage{amssymb}
\usepackage{mathtools}
\usepackage{esint}
\usepackage{pifont}
\usepackage{bbold}

\usepackage{feynmf}

\usepackage{bbm}
\usepackage{verbatim}   
\usepackage{subfigure}
\usepackage{caption}
\usepackage{acronym}
\usepackage[export]{adjustbox}

\usepackage{amsfonts}
\usepackage{amssymb}
\usepackage{mathrsfs}
\usepackage{graphicx}
\usepackage{multirow}
\usepackage{slashed}

\usepackage{caption}

\DeclareMathOperator{\Tr}{tr}
\DeclareMathOperator{\Imag}{Im}

\newcommand{\cm}{\mathcal{M}}
\newcommand{\cl}{\mathcal{L}}
\newcommand{\co}{\mathcal{O}}
\newcommand{\crot}{\mathcal{R}}
\newcommand{\I}{\mathbb{1}_3}
\newcommand{\mm}{\mathbb{M}}

\newcommand{\sw}{{\rm s}_w}
\newcommand{\cw}{{\rm c}_w}
\newcommand{\tw}{{\rm t}_w}

\definecolor{mypurple}{RGB}{164,64,214}

\title{P not PQ}

\author[a]{Nathaniel Craig,}
\emailAdd{ncraig@physics.ucsb.edu}
\author[b]{Isabel Garcia Garcia,}
\emailAdd{isabel@kitp.ucsb.edu}
\author[a]{Giacomo Koszegi,}
\emailAdd{koszegi@physics.ucsb.edu}
\author[a]{and Amara McCune}
\emailAdd{amara@physics.ucsb.edu}

\affiliation[a]{Department of Physics, University of California, Santa Barbara, CA 93106, USA}
\affiliation[b]{Kavli Institute for Theoretical Physics, University of California, Santa Barbara, CA 93106, USA}

\abstract{Parity solutions to the strong CP problem are a compelling alternative to approaches based on Peccei-Quinn symmetry, particularly given the expected violation of global symmetries in a theory of quantum gravity. The most natural of these solutions break parity at a low scale, giving rise to a host of experimentally accessible signals. We assess the status of the simplest parity-based solution in light of LHC data and flavor constraints, highlighting the prospects for near-future tests at colliders, tabletop experiments, and gravitational wave observatories. The origin of parity breaking and associated gravitational effects play crucial roles, providing new avenues for discovery through EDMs and gravity waves.
These experimental opportunities underline the promise of generalized parity, rather than Peccei-Quinn symmetry, as a robust and testable solution to the strong CP problem.
}

\begin{document}

\maketitle

\section{Introduction}
\label{sec:intro}

The current upper bound on the size of the neutron electric dipole moment (EDM) is $|d_n| < 1.8 \cdot 10^{-26} \ e \cdot {\rm cm}$ \cite{Abel:2020gbr}.\footnote{Here we quote the direct limit; the inferred bound $|d_n| < 1.6 \cdot 10^{-26} \ e \cdot {\rm cm}$ from the $^{199}$Hg EDM limit \cite{Graner:2016ses} is comparable assuming no additional contributions to the atomic EDM.} In turn, this severely constrains the size of the QCD vacuum angle, which is required to be
\begin{equation}
	\bar \theta = \theta_s + \theta_q \lesssim 10^{-10} \ ,
\label{eq:thetabound}
\end{equation}
where $\theta_q $ is the argument of the determinant of the quark mass matrix, and $\theta_s$ the coefficient of the $G \tilde G$ operator,
\begin{equation}
	\cl \supset \frac{ \theta_s \alpha_s}{4 \pi} {\rm tr} \left( G^a \tilde G^a \right) \ .
\end{equation}
In the Standard Model (SM), $\theta_q = {\rm arg} \ {\rm det} (y_u y_d) $, with $y_{u,d}$ the Yukawa matrices in the up- and down-quark sectors.
 $\bar \theta$ provides a physical, basis-independent measurement of $CP$-violation in the strong sector of the SM.

That $\bar \theta$ is constrained to be so tiny is one of the most puzzling features of the SM, and it is known as the strong CP problem. It stands alongside the electroweak hierarchy problem and the cosmological constant problem as one of the three great naturalness puzzles that remain unsolved. Although numerically the strong CP problem is orders of magnitude less severe than either of its siblings, it is considerably more robust against anthropic arguments.\footnote{See, however, \cite{Kaloper:2017fsa} for arguments to the contrary.} As such, it has drawn renewed attention during an era in which LHC null results are challenging naturalness-based approaches to the electroweak hierarchy.

Although one could argue that $\theta_s = 0$ on the basis that QCD interactions otherwise preserve $CP$, a similar argument cannot be made for a vanishing $\theta_q$. For example, if $CP$ were a good symmetry of the Yukawa sector then the Yukawa matrices would need to be real. However, real Yukawas would lead to a vanishing phase in the CKM matrix, in direct conflict with the $\co(1)$ $CP$-violation observed in the electroweak sector of the SM.
Besides $CP$, a non-zero $\bar \theta$ also violates $P$. Again, the fact that the strong sector preserves parity may allow us to impose $\theta_s = 0$. $P$ invariance in the Yukawa sector would require the Yukawa matrices to be hermitian, in which case $\theta_q = 0$ too, while still allowing for a non-zero CKM phase. However, the fact that $P$ is maximally violated by the electroweak interactions severely weakens this line of reasoning as an attempt to argue for a small $\bar \theta$.

So although $\bar \theta$ is a measurement of both $P$ and $CP$ violation by strong dynamics, the above discussion highlights how the origin of the strong CP problem in the SM ultimately lies in the features of the \emph{electroweak} sector.
It is the fact that electroweak interactions maximally violate both $P$ and $CP$ that precludes an understanding of the bound in Eq.(\ref{eq:thetabound}) based on the underlying symmetries of the SM.

With this in mind, it is natural to attempt an understanding of the smallness of $\bar \theta$ in the context of theories with an extended electroweak sector. If either $P$ or $CP$ are good symmetries of the extended theory, then $\bar \theta$ will be forced to vanish. Of course, to account for the $P$ and $CP$ violation we observe in nature, they must eventually be broken, and a non-zero $\bar \theta$ will be radiatively generated. If the induced $\bar \theta$ is small enough, this class of theories offer a symmetry-based solution to the strong CP problem. Concrete implementations of this idea based on spontaneously broken $P$ and $CP$ were first proposed in \cite{Babu:1988mw,Babu:1989rb,Barr:1991qx} and \cite{Nelson:1983zb,Barr:1984qx} respectively. It is the former that will be the focus of this work.\footnote{For recent exploration along these lines, see also \cite{Chakdar:2013tca, DAgnolo:2015uqq, Hall:2018let}.}

There is another good reason to consider solutions to strong CP based on the restoration of spacetime symmetries, namely that these may be realized as gauge symmetries in the context of string theory \cite{Dine:1992ya,Choi:1992xp}. As such, they can only be broken spontaneously and not explicitly. Depending on the scale of spontaneous symmetry breaking, the apparent lack of $P$ and $CP$ violation in the strong sector could therefore be fully, or partially, explained in this context. Clearly, a resolution to the strong CP problem along these lines would be especially attractive: it would allow us to understand the smallness of $\bar \theta$ as an accident resulting from the underlying structure of the UV-completion, as opposed to being the result of a model-building effort specially designed to address Eq.(\ref{eq:thetabound}).

From the bottom-up, there are a number of ways the SM can be extended to accommodate spontaneously broken $P$. However, in order to address the strong CP problem, a necessary feature of all of them is the presence of an $SU(2)_R$ gauge factor, as well as an extended matter content that mirrors that of the SM. Crucially, the $SU(3)$ quantum numbers of SM fermions and their mirror counterparts must be the same, making the presence of additional colored particles an irreducible feature of these models. With this extended field content, parity enforces the Yukawa couplings in the two sectors to be identical. To be phenomenologically acceptable, parity must be broken at some scale $v'$ above the weak scale, with the additional gauge bosons and mirror quarks being sufficiently heavy to evade experimental constraints. Na\"ively, bounds on the mass of colored particles would seemingly require $y_u v ' \gtrsim 1 \ {\rm TeV}$ \cite{Aaboud:2018pii,Sirunyan:2018omb}, in turn setting a lower bound $v' \gtrsim 10^8 \ {\rm GeV}$. 
But a parametric separation of scales between $v$ and $v'$ entails an irreducible amount of fine-tuning $\Delta^{-1} \simeq 2 v^2 / v'^2$, which would become $\Delta^{-1} \lesssim 10^{-12}$ for such a stringent bound on $v'$. Considering that the goal is to naturally explain a number of $\co (10^{-10})$, parity would hardly seem to remain an attractive solution to strong CP.

In this paper, we show that the conclusion of the previous paragraph is premature, and that a parity-breaking scale as low as $18 \ {\rm TeV}$ is consistent with all experimental constraints. This significantly improves the level of fine-tuning, and leaves an open window for symmetry-based solutions to strong CP that are based on spontaneously broken parity. The leading constraint on the low-tuning version of these models comes not from bounds on colored particles, but from direct searches for $Z'$ and $W'$ resonances at the LHC \cite{Aad:2019fac,Aad:2019wvl}. Future searches for heavy gauge bosons at current and future colliders are the most promising probes of this class of theories, with a $100 \ {\rm TeV}$ proton collider guaranteed to make a discovery if the level of fine-tuning is better than $\Delta^{-1} \sim 10^{-5}$ \cite{Helsens:2642473, Abada:2019lih}. Overall, the viability of these parity-based models makes collider experiments a central testing ground for solutions to strong CP.

Another attractive feature of this class of solutions to the strong CP problem is that they are robust against the effects of symmetry-breaking higher dimensional operators (HDOs) that may arise from short-distance physics associated with a gravitational UV completion. If parity is a gauge symmetry of the underlying theory, we are led to consider only those HDOs proportional to the source of spontaneous symmetry breaking. On the other hand, if parity were global, the expectation that quantum gravity violates all global symmetries \cite{Zeldovich:1976vq,Zeldovich:1977be,Banks:1988yz,Giddings:1987cg,Lee:1988ge,Abbott:1989jw,Coleman:1989zu,Kallosh:1995hi,Banks:2010zn,Harlow:2018tng,Harlow:2018jwu,Fichet:2019ugl,Daus:2020vtf} suggests we should include all HDOs that explicitly violate $P$. Although the nature of the operators is different in the gauge and global implementations, the conclusion will be the same: in both cases, the leading HDOs with $\co(1)$ coefficients may be present without destabilizing the solution to strong CP.

This stands in stark contrast with the reality of what has traditionally been the most popular solution to the strong CP problem: the QCD axion \cite{Peccei:1977hh,Peccei:1977ur,Wilczek:1977pj,Weinberg:1977ma,Kim:1979if,Shifman:1979if,Dine:1981rt}. In this case, the parameter $\bar \theta$ is promoted to the status of dynamical field, the axion, which is a pseudo-Nambu-Goldstone boson of a spontaneously broken $U(1)_{PQ}$ \emph{global} symmetry. A potential for the axion is induced non-perturbatively by QCD dynamics, and its vacuum expectation value (vev) adjusts such that $\bar \theta = 0$, thereby solving strong CP. To work, the QCD axion potential must dominate to 1 part in $10^{10}$, overwhelming any other contributions that may arise from additional degrees of freedom. New dynamics responsible for, say, dark matter, baryogenesis, or addressing the hierarchy problem, cannot significantly contribute to the axion potential. Similarly, Planck-suppressed HDOs that break $U(1)_{PQ}$ must be exceptionally suppressed \cite{Barr:1992qq,Kamionkowski:1992mf,Holman:1992us,Ghigna:1992iv}. The mechanism is not robust. The need for $U(1)_{PQ}$ to be a high quality global symmetry has become known as the ``axion quality problem''. Attempts to turn the QCD axion into a high quality axion are valuable \cite{Chun:1992bn,Randall:1992ut,Cheng:2001ys,Arvanitaki:2009fg,Fukuda:2017ylt,DiLuzio:2017tjx,Lillard:2017cwx,Lillard:2018fdt}, but hardly helpful in making a small $\bar \theta$ appear natural.

The goal of this work is to identify the most natural parity-based solution to the strong CP problem, and highlight its experimental consequences. We do so by following a strategy that combines the traditional notion of naturalness with the expectation that gravity violates all global symmetries. The former singles out a specific implementation of the spectrum of parity-symmetric models, and underscores the central role of collider experiments in exploring solutions to strong CP. The latter opens up an entirely new avenue of exploration for parity solutions to the strong CP problem, ranging from EDM experiments to gravitational wave observatories, depending on the degree to which the symmetry remains approximate.

To this end, this article is organized as follows. In section~\ref{sec:parity} we review the main features of parity-based solutions to strong CP, and discuss how a low symmetry breaking scale can be realized while complying with experimental constraints. We focus on the main phenomenological signatures of these models that are relevant for collider and flavor experiments in section~\ref{sec:pheno}. In section~\ref{sec:thetabar}, we discuss the size of radiative corrections to both $\bar \theta$ and the EDM of elementary fermions, including charged leptons, depending on the details of the parity-breaking sector. We explore the effect of Planck-suppressed HDOs on this class of models in section~\ref{sec:HDO}, paying special attention to a potential gravitational wave signal from the spontaneous breaking of parity. Section~\ref{sec:conclusions} contains our conclusions. Finally, a series of appendices contain results that have been crucial in our analysis, but may be skipped on a first reading of the manuscript.

\section{$P$ to solve strong CP}
\label{sec:parity}

In this section, we introduce the main features of symmetry-based solutions to the strong CP problem based on parity. In \ref{sec:basicidea} we review the basic idea, as first introduced in \cite{Babu:1988mw,Babu:1989rb,Barr:1991qx}. We focus on the scalar potential in \ref{sec:scalarsector}, with an emphasis on the implications for fine-tuning of the weak scale that arise as a result of the breaking of parity. In \ref{sec:fermionmasses} we discuss how the scale of additional colored particles can be decoupled from the parity-breaking scale, in turn minimizing the level of fine-tuning.

\subsection{Parity as a solution to the strong CP problem}
\label{sec:basicidea}

A symmetry-based solution to the strong CP problem based on parity requires extending the SM both in terms of matter content and gauge interactions. The minimal implementation of this idea is based on the gauge group
\begin{equation}
	SU(3) \times SU(2)_L \times SU(2)_R \times U(1)_{\hat Y} ,
\label{eq:gaugegroup}
\end{equation}
as well as a doubling of the matter content of the SM into a `mirror' sector with identical quantum numbers, except that $SU(2)_L$ doublets are now doublets of $SU(2)_R$.
Table~\ref{tab:model} summarizes the gauge charges in the quark and Higgs sectors of the theory. (Analogous charge assignments apply in the lepton sector, which we don't make explicit.) Crucially, the Higgs sector of the theory does not introduce additional sources of $CP$-violation --- indeed, the freedom to perform both $SU(2)_L$ and $SU(2)_R$ gauge transformations allows us to expand around a vacuum where both vevs are real.
\begin{table}[h]
  \begin{center}
    \begin{tabular}{l|c c c c | c c c c} 
          & $ Q = \begin{pmatrix} u  \\ d \end{pmatrix} $ & $U^\dagger$ & $D^\dagger$ & $H$
          & $ Q'^\dagger = \begin{pmatrix} u'^\dagger  \\ d'^\dagger \end{pmatrix} $ & $U'$ & $D'$	& $H'^*$ \\
      \hline
      $SU(3)$			& $\mathbf{3}$		& $\mathbf{3}$		& $\mathbf{3}$		& $\mathbf{\cdot}$
      					& $\mathbf{3}$		& $\mathbf{3}$ 		& $\mathbf{3}$		& $\mathbf{\cdot}$\\
      $SU(2)_L $		& $\mathbf{2}$		& $\mathbf{\cdot}$ 	& $\mathbf{\cdot}$	& $\mathbf{2}$
      					& $\mathbf{\cdot}$	& $\mathbf{\cdot}$	& $\mathbf{\cdot}$ 	& $\mathbf{\cdot}$\\
      $SU(2)_R $		& $\mathbf{\cdot}$	& $\mathbf{\cdot}$ 	& $\mathbf{\cdot}$	& $\mathbf{\cdot}$
      					& $\mathbf{2}$		& $\mathbf{\cdot}$	& $\mathbf{\cdot}$	& $\mathbf{2}$\\
      $U(1)_{\hat Y} $	& $\frac{1}{6}$		& $\frac{2}{3}$		& $-\frac{1}{3}$		& $\frac{1}{2}$ 
      					& $\frac{1}{6}$		& $\frac{2}{3}$		& $-\frac{1}{3}$		& $\frac{1}{2}$
    \end{tabular}
  \end{center}
  \caption{Quantum numbers in the quark and Higgs sectors. Mirror sector fields are distinguished with a prime. We use notation such that all of $Q$, $U$, and $D$ (as well as their mirror counterparts $Q'$, $U'$, and $D'$) are left-handed, two-component Weyl fermions, whereas daggered fields are always right-handed.}
      \label{tab:model}
\end{table}

With this additional field content, the theory admits an alternative definition of parity that combines the action of the `ordinary' parity transformation with an internal symmetry that exchanges the fields of the SM and mirror sectors. Explicitly, in the gauge, quark, and Higgs sectors:
\begin{align}
	{\bf W}^\mu_L & \leftrightarrow {{\bf W}_R}_\mu , \\
	Q , U, D & \leftrightarrow  Q'^\dagger , U'^\dagger, D'^\dagger , \\
	H & \leftrightarrow  H'^* ,
\end{align}
and similarly for leptons. Since $SU(3)$ and $U(1)_{\hat Y}$ interactions are not mirrored, the corresponding gauge fields transform as usual under parity. Unlike ordinary parity in the SM, this `generalized' parity transformation is now a good symmetry of the gauge sector of the theory, thanks to the extended electroweak sector and matter content.

In this context, the strong CP problem is solved as follows. On the one hand, parity requires that $\theta_s = 0$, just as one may argue in the SM based on the properties of the strong sector alone. On the other hand, the presence of additional colored particles results in an extended quark mass matrix. In particular, Yukawa terms can be written for both the SM and mirror sectors, of the form
\begin{equation}
	\cl \supset - \left\{ (y_u)_{ij} Q_i H U_j + (y'_u)_{ij} Q'^\dagger_i H'^* U'^\dagger_j \right\} + {\rm h.c.} ,
\label{eq:yukawas}
\end{equation}
and similarly for down-type quarks and leptons. As a result, the tree-level value of $\theta_q$ in these models is given by
\begin{equation}
	\theta_q = {\rm arg} \ {\rm det} (y_u y_d) + {\rm arg} \ {\rm det} (y'^*_u y'^*_d) .
\label{eq:thetaf}
\end{equation}
Crucially, demanding that Yukawa interactions preserve parity, which is now a good symmetry of the extended electroweak sector, enforces the Yukawa couplings in the two sectors to be identical, i.e.~
\begin{equation}
	y'_f = y_f .
\end{equation}
In turn, this implies $\theta_q = 0$, as per Eq.(\ref{eq:thetaf}), forcing $\bar \theta$ to vanish at tree-level in parity-symmetric models.

With the field content outlined in table~\ref{tab:model}, the theory admits an additional fermion mass term involving only the $SU(2)$-singlets, of the form
\begin{equation}
	\cl \supset - ( \cm_u )_{ij} U_i U'_j + {\rm h.c.} 
\label{eq:VLmasses}
\end{equation}
(with analogous terms for down-type quarks and leptons), where invariance under generalized parity requires that the vector-like mass matrix be hermitian, i.e.~$\cm_f^\dagger = \cm_f$.\footnote{Note that a non-hermitian mass matrix is compatible with softly broken parity; we will explore the consequences of such soft breaking in section \ref{sec:Psoft}.} In general, the expression for $\theta_q$ can be conveniently written as 
\begin{equation}
	\theta_q = {\rm arg} \ {\rm det} ( \mm_u \mm_d ) ,
\end{equation}
where $\mm_u$ and $\mm_d$ are $6 \times 6$ matrices, of the form
\begin{equation}
	\mm_f =  \begin{pmatrix} \mathbb{0} & \frac{v'}{\sqrt{2}} y'^*_f \\ \frac{v}{\sqrt{2}} y^T_f & \cm_f \end{pmatrix} , \qquad {\rm for} \qquad f=u,d  .
\label{eq:M6x6}
\end{equation}
Due to the zero in the upper-left block of the overall $6 \times 6$ mass matrix, the expression for $\theta_q$ remains as in Eq.(\ref{eq:thetaf}).
As we will discuss in \ref{sec:fermionmasses}, the presence of vector-like masses is crucial in implementing a version of the model with low fine-tuning.\footnote{A variation on the model we have so far discussed entails extending the gauge group in Eq.(\ref{eq:gaugegroup}) with an additional $U(1)$, as first discussed in \cite{Barr:1991qx}. In this case, SM and mirror fields are charged under different $U(1)$ factors, which transform into each other under parity. Although this seems like a minimal modification of the model presented here, this two-$U(1)$ version does not allow for the vector-like mass terms of Eq.(\ref{eq:VLmasses}), in turn precluding the implementation of a low parity-breaking scale.}

To obtain a phenomenologically viable model, parity must be broken, with different vevs in the mirror and SM Higgs sectors. This will induce a non-vanishing $\bar \theta$ beyond tree-level, which must be small enough if the theory is to remain a bona-fide solution to strong CP. The size of radiative corrections depends on the details of how parity is spontaneously broken. If $P$ is broken without breaking $CP$, then the radiatively induced $\bar \theta$ will be no larger than in the SM \cite{Barr:1991qx}, where $\bar \theta < 10^{-19}$ \cite{Ellis:1978hq}. On the other hand, if $CP$ is also spontaneously broken (e.g.~through the vev of a pseudo-scalar) then a larger $\bar \theta$, as well as a neutron EDM \emph{independent} of $\bar \theta$, may be radiatively generated. Even in this latter case, we will see that radiative corrections can be small enough to remain compatible with experimental constraints. Given that the final size of $\bar \theta$ is a somewhat model-dependent feature of this class of models, we defer a more detailed discussion of this issue to section~\ref{sec:thetabar}.

More generally, discussing the leading effect of broken parity on the fine-tuning of the electroweak sector does not require committing to a specific implementation of spontaneous symmetry breaking. For this purpose, it will be enough to focus on the features of the Higgs sector, to which we now turn.

\subsection{Scalar sector and fine-tuning}
\label{sec:scalarsector}

For the time being, we will parametrize the necessary breaking of parity through an explicit soft term in the scalar potential. Of course, such soft breaking should ultimately be the result of some spontaneous symmetry breaking dynamics, as we will make more explicit in section~\ref{sec:thetabar}. In this spirit, the most general scalar potential involving the $SU(2)_L$ and $SU(2)_R$ Higgs doublets takes the form
\begin{equation}
V(H,H') = - m_H^2 (|H|^2 + |H'|^2) + \lambda (|H|^2 + |H'|^2)^2 + \kappa (|H|^4 + |H'|^4) + \mu^2 |H|^2 .
\label{eq:scalarsoft}
\end{equation}
At this level, Eq.(\ref{eq:scalarsoft}) is identical to the scalar potential of theories of Neutral Naturalness, such as Twin Higgs \cite{Chacko:2005pe,Chacko:2005un}. The first two terms respect both parity and a larger accidental $SU(4)$ (really, $O(8)$) symmetry, while $\kappa$ respects the former but not the latter. The parameter $\mu^2$ softly breaks parity. In the interest of a non-trivial vacuum structure, we take $m_H^2 > 0$. Depending on the relative signs and sizes of the quartic couplings, the tree-level vacua for $\mu^2 = 0$ either preserve parity (with $v' = v$) or spontaneously break parity (with $v \neq 0, v' = 0$ or $v = 0, v' \neq 0$). A vacuum with $v' \gg v \neq 0$ may be obtained by deforming the theory away from the parity-symmetric vacuum with nonzero $\mu^2$ and $\lambda, \kappa > 0$. At tree-level, the vevs in the SM and mirror Higgs sectors are then given by
\begin{equation}
	v^2 = \frac{m_H^2 - \mu^2 ( 1 + \lambda / \kappa)}{2 \lambda + \kappa} , \qquad {\rm and} \qquad v'^2 = \frac{m_H^2 + \mu^2 \lambda / \kappa}{2 \lambda + \kappa} ,
\label{eq:vevs}
\end{equation}
where $v^2 \ll v'^2$ is necessary in order to obtain a phenomenologically viable model. After spontaneous symmetry breaking, the spectrum of the theory contains two scalar fields, $h$ and $h'$, with masses $m_h \simeq 2 \sqrt{\kappa} v$ and $m_{h'}  \simeq \sqrt{2 \lambda} v' $ respectively, as well as six Goldstones that become the longitudinal components of the gauge bosons of our extended electroweak sector. The physical gauge boson spectrum contains $Z'$ and $W'$ resonances, which are heavier than their SM counterparts by a factor of $v' / v$. We defer further details of the scalar and gauge sectors to appendix \ref{sec:appgauge}.

It is clear from Eq.(\ref{eq:vevs}) that to obtain a hierarchy of scales between $v$ and $v'$ we need to introduce a tree-level tuning between the parity-preserving and parity-breaking mass-squared terms. Heuristically, the necessary fine-tuning is given by
\begin{equation}
	\Delta^{-1} \equiv \frac{m_H^2 - \mu^2 ( 1 + \lambda / \kappa)}{m_H^2} \simeq \frac{(2 \lambda + \kappa) v^2}{(\lambda + \kappa) v'^2} \simeq \frac{2 v^2}{v'^2} .
\label{eq:tuning}
\end{equation}
This is an irreducible contribution to the fine-tuning in this class of models. Insofar as it involves the sensitivity of the weak scale $v$ to underlying parameters, it may be classified as a tuning associated with the electroweak hierarchy problem, although it is not necessarily the only such contribution. For example, a hierarchy of scales $v'^2 \ll M_{Pl}^2$ would constitute an additional source of fine-tuning in the absence of a stabilizing mechanism. Similarly, additional hierarchies of scales or couplings in the sector responsible for spontaneously breaking $P$ might necessitate similar accurate cancellations. However, these are issues that could, at least \emph{in principle}, be addressed at some higher scale above $v'$, provided the necessary dynamics do not spoil the smallness of $\bar \theta$ \cite{Albaid:2015axa}. In contrast, Eq.(\ref{eq:tuning}) is forced on us independently of the UV-completion. Although it is tempting to attach the tuning in Eq.(\ref{eq:tuning}) to the electroweak hierarchy problem and attribute it to anthropic selection (the perspective advocated in e.g.~\cite{DAgnolo:2015uqq, Hall:2018let}), this necessarily entails some favorable assumptions about the properties of an anthropic landscape. Here we prefer to render unto strong CP the things that are strong CP's, and take the irreducible tuning in Eq.(\ref{eq:tuning}) at face value as a measure of the degree to which a parity model naturally explains the small value of $\bar \theta$ without reintroducing tuning elsewhere.

With this in mind, in this paper we focus on implementations of parity solutions to strong CP where the level of fine-tuning, as parametrized in Eq.(\ref{eq:tuning}), is as mild as possible. This will concentrate our attention on a specific mechanism to generate fermion masses that in turn endows these models with characteristic phenomenology, as we discuss next.

\subsection{Fermion masses and a low parity-breaking scale}
\label{sec:fermionmasses}

Fermion mass terms arise from the Yukawa couplings of Eq.(\ref{eq:yukawas}), as well as from the vector-like mass involving the $SU(2)$-singlets. In total:
\begin{equation}
	\cl \supset - \left\{ \frac{v}{\sqrt{2}} (y_u)_{ij} u_i U_j + \frac{v'}{\sqrt{2}} (y_u)_{ij}^* u'_i U'_j + (\cm_u)_{ij} U_i U'_j \right\} + {\rm h.c.} ,
\label{eq:Lfermionmass}
\end{equation}
where we have already set $y'_u = y_u$, as mandated by generalized parity, and analogous mass terms are present both for down-type quarks and leptons.

The structure of Eq.(\ref{eq:Lfermionmass}) allows for two limiting realizations of the fermion spectrum. If the overall scale of the vector-like mass matrix is $M \ll v, v'$, then fermion masses are generated mainly through the Yukawa terms, as in the SM. In this case, mirror fermions would be an exact copy of the SM, just heavier by a factor of $v' / v$. Demanding that the lightest mirror quark is heavy enough to comply with current experimental constraints requires $m_u \times v'/v \gtrsim 1 \ {\rm TeV}$ \cite{Aaboud:2018pii,Sirunyan:2018omb}, in turn setting a lower bound $v' \gtrsim 10^8 \ {\rm GeV}$. As advertised in the Introduction, this sets the level of fine-tuning in the electroweak sector to $\Delta^{-1} \lesssim 10^{-12}$. The phenomenology of parity solutions to strong CP in this regime was discussed recently in \cite{DAgnolo:2015uqq}.

On the other hand, the limit $M \gg v, v'$, allows for a see-saw realization of the fermion spectrum, consisting of three light (SM-like) fermions, and three heavy fermions with mass of order $M$. A sufficiently high scale for the mass of additional colored particles can now be achieved by increasing $M$, not $v'$. This allows for a much lower parity-breaking scale, and therefore a much better level of fine-tuning. See-saw implementations of fermion masses, for both quarks and leptons, are discussed in \cite{Davidson:1987mh,Davidson:1987tr,Davidson:1989bx,Ranfone:1990jf}, and it was in fact in this context that a parity-based solution to the strong CP problem was first proposed \cite{Babu:1988mw,Babu:1989rb}. It is this second realization of the fermion spectrum that we concentrate on in this work.

The up-quark sector requires special consideration, since the see-saw mechanism cannot be applied to the top quark while maintaining perturbative Yukawas. So let us discuss the down-quark and lepton sectors first. (We will use notation appropriate to the down-quark sector, but emphasize that the same results apply for leptons.) To leading order in both $v/M$ and $v'/M$, the masses of the light and heavy fermions are obtained by diagonalizing the $3 \times 3$ hermitian matrices
\begin{equation}
	\frac{v v'}{2} y_d^* \cm_d^{-1} y_d^T , \qquad \qquad {\rm and} \qquad \qquad \cm_d ,
\label{eq:massmatrices}
\end{equation}
respectively. We make the simplifying assumption that there are no significant hierarchies in the eigenvalues of $\cm_d$, and therefore the heavy quarks appear at a common scale $\sim M$. Parametrically, light quark masses are then of the form $m_{d_i} \sim |y|^2 v v' /M$. The see-saw mechanism generates fermion masses $m_{d_i} \ll v$ while allowing for much larger Yukawa couplings than in the SM, which is obviously one of the main attractions of this class of models. Generating the $b$ quark mass through the see-saw mechanism while maintaining perturbativity sets an upper bound on the ratio $M/v'$, parametrically:
\begin{equation}
	m_b \sim |y|^2 \frac{v v'}{M} \lesssim \frac{v v'}{M} \qquad \Rightarrow \qquad \frac{M}{v'} \lesssim \frac{v}{m_b} \sim 10^2 .
\label{eq:Mmax}
\end{equation}

Rotating from the flavor to the mass eigenbasis in the fermion sector can be conveniently performed step by step at each order in perturbation theory, and we present a detailed discussion of this procedure in appendix \ref{sec:appfermions}. At zeroth order in $v^{(\prime)} / M$, it is necessary to perform unitary transformations acting separately on the $SU(2)$-singlet and doublet fields, of the form:
\begin{equation}
	d \rightarrow \co^\dagger_d d \, , \ d' \rightarrow \co^T_d d' \ , \qquad {\rm and} \qquad D' \rightarrow \co^\dagger_{D'} D' \, , \ D \rightarrow \co^T_{D'} D \, .
\label{eq:fermionrot0}
\end{equation}
$\co_d$ and $\co_{D'}$ are $3 \times 3$ unitary matrices acting on flavor space that diagonalize the first and second matrices of Eq.(\ref{eq:massmatrices}), respectively.
At first order in $v^{(\prime)} / M$, a further rotation is required that mixes the $SU(2)$-singlet and doublet fields as follows:
\begin{equation}
	 \begin{pmatrix} d \\ D' \end{pmatrix} \rightarrow \begin{pmatrix} \I & \epsilon_d^\dagger \\ -\epsilon_d & \I \end{pmatrix} \begin{pmatrix} d \\ D' \end{pmatrix} , 
	 \qquad {\rm and} \qquad
	  \begin{pmatrix} d' \\ D \end{pmatrix} \rightarrow \begin{pmatrix} \I & \epsilon'^\dagger_d  \\ - \epsilon'_d & \I \end{pmatrix} \begin{pmatrix} d' \\ D \end{pmatrix} ,
\label{eq:fermionrot1}
\end{equation}
where $\epsilon_d$ and $\epsilon'_d$ are $3 \times 3$ matrices with entries of $\co(v/M)$ and $\co(v'/M)$ respectively, and whose explicit expressions are given in Eq.(\ref{eq:epsilond}).

Using Dirac notation, the left- and right-handed components of the light and heavy mass eigenstates are then given by
\begin{equation}
	{d_i}_L = \begin{pmatrix} d_i \\ 0 \end{pmatrix} , \ {d_i}_R = \begin{pmatrix} 0 \\ d'^\dagger_i \end{pmatrix} , \qquad {\rm and} \qquad
	{D_i}_L = \begin{pmatrix} D'_i \\ 0 \end{pmatrix} , \ {D_i}_R = \begin{pmatrix} 0 \\ D^\dagger_i \end{pmatrix} .
\label{eq:Dirac}
\end{equation}
In particular, notice that the right-handed components of the light (SM-like) fermions consist of the corresponding component of the $SU(2)_R$-doublets, up to corrections of $\co(v'/M)$. This feature plays a crucial role in the phenomenology of these models. In particular, it leads to unsuppressed couplings between $SU(2)_R$ gauge bosons, and the right-handed currents of the SM-like fermions. As we will discuss in \ref{sec:irreducible}, this leads to the most stringent bound on the parity-breaking scale.

As far as the up-quark sector is concerned, the see-saw mechanism can be implemented for the $u$ and $c$ quarks, with the corresponding heavy partners appearing at the scale $\sim M$. The mass eigenstates for the first two generations are as in Eq.(\ref{eq:Dirac}). The top sector, on the other hand, cannot be significantly ``see-sawed''. Instead, it consists of light and heavy top partners with tree-level masses $m_t \simeq y_t v / \sqrt{2}$ and $m_{t'} \simeq m_t \times v' / v $, respectively. In Dirac notation, and at zeroth order in $v^{(\prime)} / M$, the mass eigenstates are now purely made of SM and mirror sector fields, i.e.~
\begin{equation}
	t_L = \begin{pmatrix} u_3 \\ 0 \end{pmatrix} , \ t_R = \begin{pmatrix} 0 \\ U_3^\dagger \end{pmatrix} , \qquad {\rm and} \qquad
	t'_L = \begin{pmatrix} U_3' \\ 0 \end{pmatrix} , \ t'_R = \begin{pmatrix} 0 \\ u_3'^\dagger \end{pmatrix} .
\label{eq:Diractop}
\end{equation}
As usual, rotation matrices in the quark sector are constrained by the requirement that the CKM matrix is reproduced appropriately, which in this case implies $V = \co_u \co^\dagger_d$, up to corrections of $\co(v^2/M^2)$. Further details concerning the mass diagonalization procedure in the fermion sector can be found in appendix \ref{sec:appfermions}.

\begin{figure}
  \centering
  \includegraphics[scale=1.0]{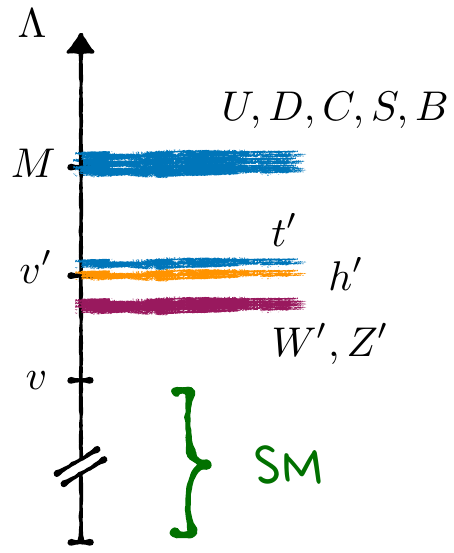}
  \caption{Schematic illustration of the particle spectrum of parity solutions to the strong CP problem in their least tuned version. The lightest exotic particles are $W'$ and $Z'$ resonances, followed by a top partner appearing at a scale of order $v'$. A mirror Higgs is also part of the low-lying spectrum, with $m_{h'} \simeq \sqrt{2 \lambda} v'$. (For illustration, we have chosen $\lambda = \co(1)$, but note that $h'$ could be much lighter if $\lambda \ll 1$.) Additional mirror quarks have masses of order the see-saw scale, $M \gg v'$. The lepton sector must also be ``see-sawed'', with mirror leptons similarly appearing well above $v'$ (although we emphasize that the see-saw scales in the quark and lepton sectors need not coincide).}
  \label{fig:spectrum}
\end{figure}

This finalizes our discussion of the main characteristics of parity solutions to strong CP that feature low fine-tuning in the electroweak sector. Before moving on, we include in figure~\ref{fig:spectrum} a schematic representation of the typical spectrum of these models. Amusingly, the combination of parity and the see-saw mechanism leads to a spectrum of partner particles strikingly reminiscent of a ``natural'' left-right Twin Higgs model \cite{Chacko:2005un,Goh:2006wj} with light top and $W/Z$ partners.

\section{Dial $P$ for Phenomenology}
\label{sec:pheno}

We now turn to the phenomenology of natural parity solutions to strong CP, beginning with direct bounds from the LHC in section \ref{sec:irreducible} before turning to indirect flavor constraints in \ref{sec:flavor}. The collider and flavor phenomenology of similar left-right models has been the topic of previous work \cite{Babu:1988mw,Babu:1989rb,Ranfone:1990jf,Kiyo:1998zm,Goh:2006wj}, and our focus here will be on those ``irreducible" signatures that are mandated by the structure of the theory in its capacity as a solution to strong CP. A more in depth study of collider and flavor signatures in light of forthcoming data can illuminate the additional structure of these models, and it is a worthwhile direction for continued study.

\subsection{Collider bounds}
\label{sec:irreducible}

The doubling of the electroweak sector gives rise to a plethora of experimental signatures at colliders, ranging from additional vector bosons (the $W'$ and $Z'$ of spontaneously broken $SU(2)_R$) to vector-like quarks (the $SU(2)$-singlet fermions) to additional Higgs bosons. Ultimately, given that the $Z'$ and $W'$ gauge bosons acquire masses exclusively from $SU(2)_R$ breaking, and inherit couplings to the SM-like quarks and leptons, collider searches for these additional vectors place the most solid and strongest direct bounds on the models under consideration.

\subsubsection*{Neutral currents}

The $Z'$ resonance inherits couplings to both the left- and right-handed currents of SM-like fermions. In the down-type quark and lepton sectors, these are both flavor diagonal and generation universal. After rotating to the mass eigenbasis in both the gauge and fermion sectors, as outlined in appendix \ref{sec:massbasis}, we find:
\begin{equation}
	\cl \supset g Z'_\mu \sum_{i=1}^3 \left(	z'_{d_R} \bar {d_i}_R \gamma^\mu {d_i}_R + z'_{e_R} \bar {e_i}_R \gamma^\mu {e_i}_R + z'_{\nu_R} \bar {\nu_i}_R \gamma^\mu {\nu_i}_R \right) + \{ R \rightarrow L \} \ .
\end{equation}
As discussed in section \ref{sec:fermionmasses}, the see-saw implementation of fermion masses leads to unsuppressed couplings between the $SU(2)_R$ gauge bosons and right-handed fermions. Up to corrections of $\co (\sin^2 \theta_w)$, these are identical to the couplings between the SM $Z$ and left-handed currents. Specifically:
\begin{equation}
	z'_{d_R} \simeq - \frac{g}{2} \left( 1 + \co(\sw^2) \right) , \quad z'_{e_R} \simeq - \frac{g}{2} \left( 1 + \co(\sw^2) \right) , \quad {\rm and} \quad z'_{\nu_R} \simeq \frac{g}{2} \left( 1 + \co(\sw^2) \right) ,
\end{equation}
where $\sw \equiv \sin \theta_w$, and we have ignored corrections of $\co (v^2 / v'^2)$. On the other hand, couplings of the $Z'$ to left-handed currents are now suppressed:
\begin{equation}
	z'_{d_L} \simeq - \frac{g}{6} \frac{ \sw \tw }{ \sqrt{\cos 2\theta_w} } = \co(\sw^2) , \quad {\rm and} \quad z'_{e_L} = z'_{\nu_L} = -3 z'_{d_L} .
\end{equation}
The situation in the up-quark sector is somewhat different. Now, couplings between the $Z'$ and the right-handed fermion currents are no longer universal. Instead, we have:
\begin{equation}
	\cl \supset g Z'_\mu \left(	\sum_{i=1}^3 z'_{u_L} \bar{u_i}_L \gamma^\mu {u_i}_L + \sum_{i=1}^2 z'_{u_R} \bar {u_i}_R \gamma^\mu {u_i}_R + z'_{t_R} \bar t_R \gamma^\mu t_R \right) .
\end{equation}
As before, $Z'$ couplings to first and second generation right-handed currents are unsuppressed, and are given by
\begin{equation}
	z'_{u_R} \simeq \frac{g}{2} \left( 1 + \co(\sw^2) \right) \ ,
\end{equation}
whereas those to left-handed fermions, as well as to the right-handed top, now read
\begin{equation}
	z'_{u_L} = \frac{z'_{t_R}}{4} \simeq - \frac{g}{6} \frac{ \sw \tw }{ \sqrt{\cos 2\theta_w} } = \co (\sw^2) \ .
\end{equation}

Bounds on the $Z'$ mass from its production at the LHC will therefore be similar to those found in the so-called Sequential Standard Model, which features a $Z'$ resonance that is just a heavy copy of the SM $Z$. In the present model, couplings of the $Z'$ to SM fermions are similar to those of the $Z$ after the replacement $L \leftrightarrow R$ --- a replacement that does not affect either the production cross section or the decay rates into light fermions. The most constraining limits thus come from \cite{Aad:2019fac}, where a search focused on leptonic final states sets a lower bound $m_{Z'} \gtrsim 5$~TeV. In turn, this translates into a lower limit on the scale of parity breaking of order $v' \gtrsim 13$~TeV.

\subsubsection*{Charged currents}

$W'$ gauge bosons interact with the right-handed SM fermions in a way that mirrors the interactions between their left-handed counterparts and SM $W$. In the lepton sector:
\begin{equation}
	\cl \supset \frac{g}{\sqrt{2}} \sum_{i,j=1}^3 \left(	B_{ij} W^+_\mu \bar {\nu_i}_L \gamma^\mu {e_j}_L	+	B'_{ij} W'^+_\mu \bar {\nu_i}_R \gamma^\mu {e_j}_R	\right) + {\rm h.c.} ,
\end{equation}
where $B = B' = \co_\nu \co_e^\dagger$, up to corrections of order ${v^{(\prime)}}^2 / M^2$.
As far as the quark sector is concerned, the up-type sector again requires special consideration. We find:
\begin{equation}
	\cl \supset \frac{g}{\sqrt{2}}	\sum_{i,j=1}^3 W^+_\mu V_{ij} \bar {u_i}_L \gamma^\mu {d_j}_L + {\rm h.c.} ,
\end{equation}
with $V = \co_u \co_d^\dagger + \co(v^2/M^2)$, whereas
\begin{equation}
	\cl \supset \frac{g}{\sqrt{2}} W'^+_\mu	\sum_{j=1}^3 \left(	\sum_{i=1}^2 V'_{ij} \bar {u_i}_R \gamma^\mu {d_j}_R + \Delta V'_{3 j} \bar t_R \gamma^\mu {d_j}_R	\right) + {\rm h.c.}
\end{equation}
Up to corrections of $\co(v'^2/M^2)$, we have $V' = V$, and $\Delta V'_{3 j} = (\epsilon'^*_u V)_{3j}$. The $3 \times 3$ matrix $\epsilon'_u$, whose entries are suppressed by a factor of $v'/M \ll 1$, is given explicitly in Eq.(\ref{eq:epsilonu}).

As with the $Z'$, we expect bounds on the $W'$ to be comparable to those in the Sequential Standard Model. Current direct searches set stringent constraints on such $W'$ resonances, of order $m_{W'} \gtrsim 6$ TeV \cite{Aad:2019wvl}. In turn, this sets the strongest limit on the scale of parity breaking: $v' \gtrsim 18$ TeV.  Although direct searches for vector-like quarks and additional Higgs bosons are also germane, they lead to significantly weaker bounds on the scale of parity breaking compared to $W'$ and $Z'$ searches. For example, null results in searches for vector-like top partners \cite{Aaboud:2018pii,Sirunyan:2018omb} lead to $v' \gtrsim 2$ TeV, with comparable bounds coming from searches for SM-singlet scalars. \\

Looking to the future, a 100 TeV $pp$ collider such as the proposed FCC-hh should be sensitive to $W'$ and $Z'$ bosons as heavy as $\sim 40$ TeV \cite{Helsens:2642473, Abada:2019lih}, corresponding to $v' \gtrsim 120$ TeV. This would comprehensively cover the most natural parameter space consistent with current data, and the non-observation of heavy vectors at such a collider would suggest that parity solutions are tuned at the $\Delta^{-1} \sim 10^{-5}$ level. In this respect, future colliders provide a decisive test of parity solutions to the strong CP problem.

\subsection{Flavor constraints}
\label{sec:flavor}

In the SM, flavor-changing neutral currents (FCNCs) are absent at tree-level, appearing only at one-loop, and being additionally suppressed by the GIM mechanism. As a result, precision measurements of flavor-violating processes often imply stringent constraints on extensions of the SM. In the class of models under consideration, FCNCs arise already at tree-level, mediated by the $Z$ and $Z'$ gauge bosons, as well as the scalars $h$ and $h'$. However, their size is suppressed by factors of the Yukawa couplings of the relevant fermions, making their effect negligible.
At one-loop, FCNCs proceeding via box diagrams involving $W'$ gauge bosons and mirror up-type quarks can lead to deviations in kaon properties, in turn setting the leading constraints on the flavor structure of these models.

\subsubsection*{Tree-level FCNCs}

Rotating from the gauge to the mass eigenbasis in the fermion and gauge boson sectors, as specified in appendix \ref{sec:massbasis}, leads to the presence of flavor-changing interactions between the $Z$ and the SM-like fermions. For example, in the down-quark sector there are new interactions of the form
\begin{equation}
	\cl \supset \frac{g}{2 \cw} (\epsilon_d^\dagger \epsilon_d)_{ij} Z_\mu \bar {d_i}_L \gamma^\mu {d_j}_L ,
\end{equation}
where $\epsilon_d$ is a $3 \times 3$ matrix acting on flavor space whose explicit form is given in Eq.(\ref{eq:epsilond}).
Integrating out the $Z$, the effective hamiltonian relevant to describe $| \Delta F | = 1$ processes, such as the leptonic decay of $B$ mesons, now contains additional terms, of the form
\begin{equation}
	\Delta \mathcal{H}_{\rm eff} \simeq - \sqrt{2} G_F (\epsilon_d^\dagger \epsilon_d)_{32} \cos (2 \theta_w) (\bar b_L \gamma^\mu s_L) (\bar \mu_L \gamma_\mu \mu_L) + {\rm h.c.}
\label{eq:Heff}
\end{equation}
(An analogous term involving right-handed muons is also present, but suppressed by a factor of $\sw^2$, so we neglect it in the subsequent discussion.)

The deviation with respect to the SM prediction for the branching fraction of the process $B_s^0 \rightarrow \mu^+ \mu^-$ as a result of the operator in Eq.(\ref{eq:Heff}) can be written as
\begin{equation}
	r_{\mu \mu} \equiv \frac{ {\rm BR} (B^0_s \rightarrow \mu^+ \mu^-)_{\rm BSM} }{ {\rm BR} (B^0_s \rightarrow \mu^+ \mu^-)_{\rm SM} } - 1
	\simeq \frac{ | C^{\rm (SM)}_{10} + C^{\rm (BSM)}_{10} |^2}{ | C^{\rm (SM)}_{10} |^2 } - 1 ,
\label{eq:rmumu}
\end{equation}
where $C^{\rm (SM)}_{10}$ and $C^{\rm (BSM)}_{10}$ are the SM and BSM contributions to the Wilson coefficient of the four-fermion operator $(\bar b_L \gamma^\mu s_L)(\bar \mu \gamma_\mu \gamma^5 \mu)$.
In the SM
\begin{equation} 
	C^{\rm (SM)}_{10} = \frac{G_F}{2 \sqrt{2}} \frac{\alpha}{4 \pi} (V_{tb}^* V_{ts}) \tilde C^{\rm (SM)}_{10} ,
\end{equation}
with $\tilde C^{\rm (SM)}_{10} \simeq 4.41$ \cite{Khodjamirian:2010vf}, whereas from Eq.(\ref{eq:Heff}) we have
\begin{equation}
	C^{\rm (BSM)}_{10} \simeq \frac{G_F}{\sqrt{2}} \cos (2 \theta_w) (\epsilon_d^\dagger \epsilon_d)_{32} .
\end{equation}
A stringent upper bound on the size of the $(\epsilon_d^\dagger \epsilon_d)_{32}$ coefficient arises from the requirement that the masses of the down-type quarks are correctly reproduced in this model. From Eq.(\ref{eq:epsilond}), we have
\begin{equation}
	(\epsilon_d^\dagger \epsilon_d)_{32} = \frac{v^2}{2} \sum_i \frac{(\tilde y_d)_{3i} (\tilde y_d)_{2i}^*}{m_{D_i}^2}
	\sim \frac{v^2}{M^2} \sum_i (\tilde y_d)_{3i} (\tilde y_d)_{2i}^*
	\lesssim \frac{v}{M} \frac{\sqrt{m_b m_s}}{v'} \ll 1,
\end{equation}
where in the last step we have made use of the upper bound in Eq.(\ref{eq:ymax}). We then have, parametrically
\begin{equation}
	r_{\mu \mu} \sim \frac{2 |C^{\rm (BSM)}_{10}|}{|C^{\rm (SM)}_{10}|} \lesssim 10^{-3} \left( \frac{18 \ {\rm TeV}}{v'} \right)^2 \left( \frac{v'}{M} \right) .
\end{equation}
This effect is much smaller than the theoretical and experimental errors on ${\rm BR} (B^0_s \rightarrow \mu^+ \mu^-)$, which are both on the order of $10 \%$ \cite{Zyla:2020zbs}.\footnote{The effective operator $(\bar b_R \gamma^\mu s_R)(\bar \mu \gamma_\mu \gamma^5 \mu)$ is also generated after integrating out the $Z$, with a Wilson coefficient $C'^{\rm (BSM)}_{10} \sim C^{\rm (BSM)}_{10}$ that enters into Eq.(\ref{eq:rmumu}) in a similar manner. The presence of this operator does not quantitatively affect our analysis.}

The effects of $Z$-mediated FCNCs on other processes are even more suppressed. For example, $|\Delta F|=2$ processes such as kaon mixing require two insertions of the (tiny) flavor-violating coefficient. In the lepton sector, even $|\Delta F| =1$ decays are virtually unobservable, as the effect is now suppressed by the masses of the relevant leptons. FCNCs mediated by the SM Higgs are similarly negligible, since the corresponding Wilson coefficients feature the same suppression as those from $Z$ exchange, on top the smaller coupling between the Higgs and light fermions. Flavor-changing interactions mediated by the $Z'$ and $h'$ are further suppressed by an additional factor of $m_Z^2 / m_{Z'}^2$ and $m_h^2 / m_{h'}^2$ respectively, making them irrelevant. Overall, the strong suppression of the tree-level FCNCs that occurs naturally in these models makes their effects negligible.

\subsubsection*{One-loop FCNCs}

\begin{figure}
  \centering
  \includegraphics[scale=0.9]{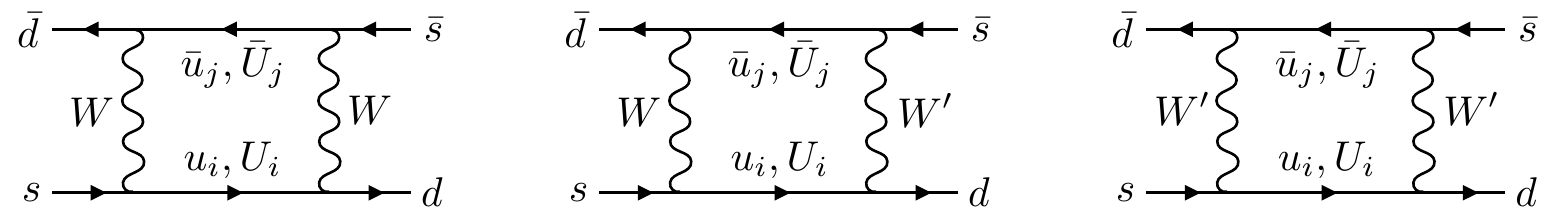}
  \caption{Additional box diagrams contributing to kaon mixing in the models under consideration include any of the up-type quarks propagating inside the loop, as well as (left) two $W$ bosons, (center) one $W$ and one $W'$, and (right) two $W'$s. Quantitatively, it is diagrams with one $W$ and one $W'$ that are the most relevant.}
  \label{fig:box}
\end{figure}

Another source of FCNCs beyond those present in the SM arises at one loop. In these models, the familiar box diagram that describes meson mixing in the SM is now accompanied by similar diagrams that include $W'$s as well as the additional (heavy) up-type quarks running inside the loop, as we show in figure \ref{fig:box}.
Given the level of experimental precision in measurements of kaon mixing parameters, even a modification to this process at the loop level can be a significant source of constraints.

The relevant interactions are those involving the SM-like down-type quarks and both the $W$ and $W'$ gauge bosons. They are given by
\begin{equation}
	\cl \supset \frac{g}{\sqrt{2}} W^+_\mu	\sum_{j,i=1}^3 \left( V_{ij} \bar {u_i}_L \gamma^\mu {d_j}_L + \Delta V_{i j} \bar {U_i}_L \gamma^\mu {d_j}_L	\right) + {\rm h.c.} ,
\label{eq:LWups}
\end{equation}
where ${U_3}_L = t'_L$ here and $\Delta V = \epsilon_u V$, as well as
\begin{equation} \begin{split}
	\cl \supset \frac{g}{\sqrt{2}} W'^+_\mu	\sum_{j=1}^3	& \left(	\sum_{i=1}^2 V_{ij} \bar {u_i}_R \gamma^\mu {d_j}_R + V_{3j} \bar t'_R \gamma^\mu {d_j}_R \right. \\
												&\left. + \sum_{i=1}^2 \Delta V'_{i j} \bar {U_i}_R \gamma^\mu {d_j}_R + \Delta V'_{3 j} \bar t_R \gamma^\mu {d_j}_R \right) ,
\label{eq:LWpups}
\end{split} \end{equation}
with $\Delta V' = \epsilon'^*_u V$. The entries of the $3 \times 3$ matrices $\epsilon_u$ and $\epsilon'_u$ are $\co (v/M)$ and $\co(v'/M)$ respectively, and explicit expressions can be found in Eq.(\ref{eq:epsilonu}).
 
The detailed expressions, including loop functions, relevant to estimate the contributions to the kaon mixing parameters $\Delta m_K$ and $|\epsilon_K|$ can be found in appendix~\ref{app:flavor}. Additional box diagrams including two $W$s or two $W'$s always lead to a contribution which is much smaller than that of the SM, and can therefore be neglected. The leading contribution arises from diagrams including one $W$ and one $W'$. In this case, there is an ``irreducible" contribution to both parameters (irreducible in the sense that it can only be ``turned-off" by increasing $v'$), which comes from the $u$ and $c$ quarks, whose couplings to the $W'$ gauge boson are set to be equal to those of the CKM matrix as a result of generalized parity. The size of this correction reads
\begin{equation}
	(\Delta m_K)_{u,c} \approx - 6 \cdot 10^{-16} \ {\rm GeV} \left( \frac{6 \ {\rm TeV}}{m_{W'}} \right)^2 , \qquad {\rm and} \qquad |\epsilon_K|_{u,c} \approx 7 \cdot 10^{-5} \left( \frac{6 \ {\rm TeV}}{m_{W'}} \right)^2 ,
\label{eq:kaonirred}
\end{equation}
which in both cases is an order of magnitude below the theoretical error in the corresponding SM prediction, for values of $m_{W'}$ consistent with the direct bounds discussed in \ref{sec:irreducible}.

Contributions from box diagrams that involve additional members of the up-quark sector additionally depend on the see-saw scale $M$, as well as the size of both diagonal and off-diagonal entries in the up-type Yukawa matrices. As far as $\Delta m_K$ is concerned, the leading contribution comes from diagrams where the $u$ and $c$ quarks propagate inside the loop, and so it is roughly equal to the result in Eq.(\ref{eq:kaonirred}), even for a see-saw scale $M$ that sits only slightly above $v'$. In contrast, the contribution to $|\epsilon_K|$ can be large, and it is dominated by diagrams where the $t$ quark propagates inside the loop. Choosing the individual entries in the Yukawa couplings to saturate the upper bound given in Eq.(\ref{eq:yumax}), $|\epsilon_K|$ sets a lower bound on $M$ than can range between $750$ TeV and $1000$ TeV (depending on whether the leading contribution interferes destructively or constructively with the SM result) for $v' \sim 18$ TeV. This value of $M$ sits comfortably within the upper bound $M \lesssim 10^2 v'$, which follows from the requirement of perturbative Yukawas, as discussed around Eq.(\ref{eq:Mmax}). Alternatively, even for $v' = 18$ TeV and $M = 40$ TeV, an additional suppression by a factor of $\co(0.1)$ in the off-diagonal elements of the up-type Yukawas with respect to their upper bound is enough to bring the predicted value of $|\epsilon_K|$ within the allowed range.

Overall, the class of parity solutions to the strong CP problem that we focus on in this work can comfortably satisfy existing constraints from flavor physics. Flavor-changing processes are, nevertheless, an interesting probe of the structure of these models, and a more in-depth investigation is a promising avenue for future work.

\section{Broken parity and the neutron EDM}
\label{sec:thetabar}

As we discussed in section~\ref{sec:parity}, parity-symmetric theories predict a vanishing $\bar \theta$, therefore offering a potential solution to the strong CP problem. However, the breaking of parity that is necessary for phenomenological reasons implies that, although zero at tree-level, a non-zero $\bar \theta$ may be generated radiatively. In this section, we investigate in detail the size of radiative corrections to both $\bar \theta$, and the EDM of elementary fermions. We focus on the effect of non-gravitational interactions, and leave gravitational considerations to section \ref{sec:HDO}.

The size of radiative corrections to the $\bar \theta$ parameter is a somewhat model-dependent question, as it depends on the details of how parity is broken.
For instance, we could regard generalized parity to be a global symmetry that is only broken softly by dimensionful parameters, as in Eq.(\ref{eq:scalarsoft}). More realistically, we might expect that the breaking of parity is spontaneous, and not explicit. This must certainly be the case if, for example, parity were a gauge symmetry of the UV theory. Even in this case, there are two qualitatively different options: either parity is broken without breaking $CP$ (e.g.~through a symmetry-breaking sector with two scalar fields that obtain asymmetric vevs); or both parity and $CP$ are broken simultaneously (e.g.~through the vev of a pseudo-scalar). The former situation is quantitatively similar to the global case. In the latter, however, the symmetry-breaking sector can introduce an additional source of $CP$-violation beyond that present in the SM, and a non-vanishing $\bar \theta$ can arise already at one loop.

In the remainder of this section, we discuss the three qualitatively different possibilities for the breaking of parity, with a focus on the implications for the size of radiative corrections to the neutron EDM.

\subsection{Softly broken parity}
\label{sec:Psoft}

We will first discuss the possibility of parity being broken softly, only as a result of dimensionful parameters. Performing this analysis will give us an understanding of the irreducible effects that will be present in any theory where the breaking of parity happens dynamically.

There are two potential sources of soft breaking. One corresponds to the $\mu^2$ term in the scalar potential of Eq.(\ref{eq:scalarsoft}), which splits the Higgs vevs in the SM and mirror sectors. As anticipated in the introduction, if this was the only source of parity-violation, radiative corrections to $\bar \theta$ would be no larger than in the SM \cite{Barr:1991qx}. Another potential source of soft breaking are the vector-like mass matrices of Eq.(\ref{eq:VLmasses}). Relaxing the requirement that these be hermitian introduces a soft breaking of both generalized parity and $CP$. In this case, a correction to the EDMs of elementary charged fermions (both quarks and leptons) arises already at one loop, whereas $\bar \theta$ remains zero both at the tree- and one-loop levels. In turn, this translates into a contribution to the neutron EDM \emph{independent} of $\bar \theta$.

\begin{figure}
  \centering
  \includegraphics[scale=1.0]{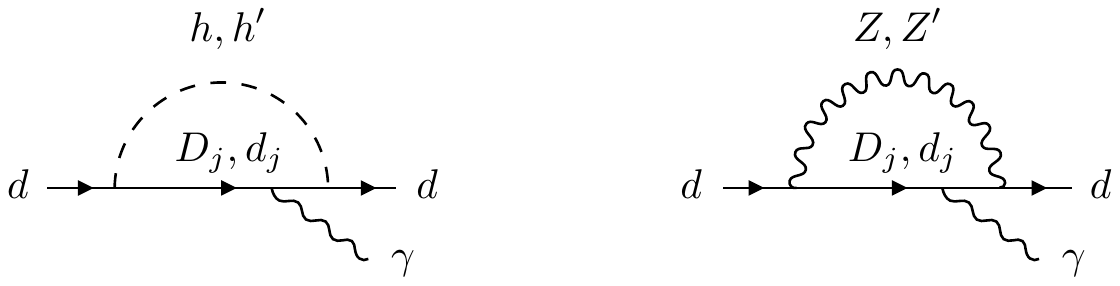}
  \caption{One-loop diagrams generating a non-zero EDM for the $d$ quark in the presence of soft breaking of generalized parity through non-hermitian vector-like mass matrices for the $SU(2)$-singlets. The leading contribution arises from the diagram where $h'$, and the heavy mirror quarks, $D_j$, propagate inside the loop. Analogous diagrams are present in both the up-quark and lepton sectors.}
  \label{fig:1loopEDM}
\end{figure}

Taking the vector-like mass matrices of the $SU(2)$-singlets to be general complex matrices, WLOG we may write them as
\begin{equation}
	\cm'_f = \cm_f + i \Delta \cm_f ,
\end{equation}
where both $\cm^\dagger_f = \cm_f$ and $\Delta \cm^\dagger_f = \Delta \cm_f$. If $\Delta \cm_f$ is non-vanishing, $\cm'_f$ is no longer hermitian, therefore (softly) breaking both generalized parity and $CP$. At one-loop, a non-zero $\Delta \cm_f$ leads to a non-vanishing contribution to the EDM of elementary fermions, with the relevant diagrams depicted in figure~\ref{fig:1loopEDM}.
The result is dominated by diagrams where the mirror Higgs, $h'$, and the heavy mirror fermions propagate inside the loop. We present a detailed calculation in appendix \ref{sec:appEDM}. For any of the light SM fermions, we find
\begin{equation}
	\frac{d_f}{e} \simeq \frac{n_f Q_f m_f}{32 \pi^2 M^2} \times \co \left( \frac{|\Delta \cm|}{M} \right) .
\end{equation}
where $n_f$ is the number of mirror fermions appearing at the see-saw scale in each fermion sector (i.e.~$n_d = n_e = 3$, and $n_u = 2$), and $|\Delta \cm|$ refers to the typical size of the entries in the $\Delta \cm$ matrix.

Taking the soft-breaking through $\Delta \cm$ to be $\co (1)$, we find, parametrically,
\begin{equation}
	d_u, d_d \sim 10^{-28} \left( \frac{40 \ {\rm TeV}}{M} \right)^2 e \cdot {\rm cm}.
\label{eq:EDMquarks}
\end{equation}
In turn, this will translate into an EDM for the neutron of approximately the same size. For illustration, we have normalized the above expression to a value of $M$ that is roughly a factor of two larger than the current lower bound on $v'$. The corresponding result lies two orders of magnitude below the current experimental bound on $d_n$, and could fall within reach of future experiments depending on the value of the see-saw scale (see e.g. \cite{Hutzler:2020lmj} for a survey of prospective molecule-based searches promising orders-of-magnitude improvement in sensitivity to hadronic CPV).

The see-saw mechanism must also be implemented in the charged lepton sector. If it were not, the mirror partner of the SM electron would appear at a scale $m_{e'} \simeq m_e \times v' /v$, which would be as low as $\sim 40$ MeV for the least fine-tuned version of the model where $v' \simeq 18$ TeV. Since mirror fermions carry the same electromagnetic charge as their SM counterparts, this possibility is obviously ruled out. As a result, a non-zero electron EDM is also a generic prediction of this class of theories. Parametrically
\begin{equation}
	d_e \sim 10^{-29} \left( \frac{90 \ {\rm TeV}}{M} \right)^2 e \cdot {\rm cm},
\end{equation}
where we have chosen the see-saw scale in the lepton sector so as to saturate the current upper bound on the electron EDM, which is $|d_e| < 1.1 \cdot 10^{-29} \ e \cdot {\rm cm}$ \cite{Andreev:2018ayy}.

Although an EDM is generated at one loop for the various elementary fermions, $\bar \theta$ remains zero at this order. At tree-level, it is easy to see that $\bar \theta = 0$, even in the presence of non-hermitian vector-like mass matrices. Working in the flavor basis, the full $6 \times 6$ mass matrices in both the up- and down-quark sectors are only modified with respect to Eq.(\ref{eq:M6x6}) by replacing $\cm_f$ with $\cm'_f$ in the bottom-right block. We then have
\begin{equation}
	\det \mathbb{M}_f = \det \begin{pmatrix} \mathbb{0}_3 & \frac{v'}{\sqrt{2}} y'^*_f \\ \frac{v}{\sqrt{2}} y^T_f & \cm'_f \end{pmatrix} \propto \det \left( y'^*_f y^T_f \right) \qquad {\rm for} \qquad f = u , d ,
\label{eq:Mp6x6}
\end{equation}
which is real regardless of $\cm'_f$, since $y'_f = y_f$. This is clearly an accidental consequence of the zero appearing in the upper-left corner of the quark mass matrix --- the gauge structure of the theory does not allow for relevant operators with the appropriate quantum numbers to fill that block. The vanishing of $\bar \theta$ at one-loop is less immediately obvious. The relevant calculation was performed in \cite{Babu:1989rb}, and it is also apparent as a byproduct of our EDM calculation in appendix \ref{sec:appEDM}. As already emphasized in \cite{Babu:1989rb}, a non-zero correction to $\bar \theta$ could appear at the two-loop order, and would lead to an additional contribution to the neutron EDM that could be comparable in size to the one discussed here.

\subsection{Spontaneously broken parity and $CP$}
\label{sec:PandCP}

Perhaps more compellingly --- and necessarily, if parity is a gauge symmetry --- the breaking of parity can be accomplished through the vacuum expectation value of an additional field. The most minimal realization actually entails the breaking of both parity and $CP$ through the vev of a pseudo-scalar field $\phi$. The soft term in Eq.(\ref{eq:scalarsoft}) is generated by pseudo-scalar couplings to the Higgs sector, of the form
\begin{equation}
	V \supset \mu_\phi \phi ( |H|^2 - |H'|^2) + \lambda_\phi \phi^2 (|H|^2 + |H'|^2) \ .
\label{eq:VphiHiggs}
\end{equation}
The first term above splits the two vevs, and to obtain $v' \gg v$ entails
\begin{equation}
	\mu_\phi v_\phi \sim \kappa v'^{2} \ .
\end{equation}
A natural possibility is to take $v_\phi \sim \mu_\phi \sim v'$, with $\kappa = \co(1)$. However, $\kappa \ll 1$ is also possible, especially since this coupling breaks the otherwise accidental $SU(4)$ global symmetry of the scalar potential in Eq.(\ref{eq:scalarsoft}). Indeed, the quartic coupling of the SM-like Higgs is $\lambda_{SM} \sim 2 \kappa,$ suggesting $\kappa \lesssim 0.1$ and thus a pseudo-scalar vev $v_\phi$ that is numerically somewhat smaller than $v'$.

Crucially, there is an additional operator consistent with all symmetries that involves $\phi$ and the $SU(2)$-singlet fermions, of the form \cite{Albaid:2015axa}:
\begin{equation}
	\cl \supset i (\bar y_d)_{i j} \phi D_i D_j' + {\rm h.c.} ,
\label{eq:yphi}
\end{equation}
and similarly for up-type quarks and leptons. The $\bar y_f$ matrices must be hermitian in order to respect generalized parity. When $\phi$ gets a vev, this term breaks both parity \emph{and} $CP$. In the notation of section \ref{sec:Psoft}, a non-hermitian contribution to the vector-like masses in the fermion sector is generated, of the form $\Delta \cm_f = \bar y_f v_\phi$. More importantly, new interactions involving the pseudo-scalar lead to a non-zero contribution to $\bar \theta$ already at one loop, which sets stringent constraints on the size of these couplings.
The relevant diagrams are those on the left of figure~\ref{fig:1loopEDM}, minus the external photon line, and allowing for $\phi$ to propagate inside the loop. In appendix~\ref{sec:apptheta}, we present a detailed calculation, performed in the mass eigenbasis, of the one-loop correction to the quark mass matrix, and the corresponding correction to $\bar \theta$, in the context of left-right models with a see-saw fermion structure. In the remainder of this section, we will reproduce the parametric contribution to $\bar \theta$ from the down-quark sector using a spurion analysis that the reader might find more instructive.

If we parametrize the one-loop correction to the $6 \times 6$ mass matrix in the down-quark sector in terms of $3 \times 3$ blocks, as follows
\begin{equation}
	\Delta \mm_d \equiv
	\begin{pmatrix}
  	\Delta_{d' d} & \Delta_{d' D'} \\
    	\Delta_{D d} & \Delta_{D D'}
    	\end{pmatrix},
\end{equation}
then the corresponding contribution to $\bar \theta$ from the down-quark sector can be written as
\begin{equation}
\begin{split}
	\theta_d 	& \equiv {\rm arg} \ \det \left( \mm_d + \Delta \mm_d \right)   \\
			& \simeq \Imag \Tr \left( \mm^{-1}_d \Delta \mm_d \right) \\
			& = \Imag \Tr \left\{ - \left( \frac{v v'}{2} y'^*_d \cm^{-1}_d y^T_d \right)^{-1} \Delta_{d' d} + \left( \frac{v}{\sqrt{2}} y^T_d \right)^{-1} \Delta_{D d} + \left( \frac{v'}{\sqrt{2}} y'^*_d \right)^{-1} \Delta_{d' D'} \right\}  .
\end{split}
\label{eq:thetadloop}
\end{equation}
Notice that the $\Delta_{D D'}$ block does not contribute to $\theta_d$ at this order, which again is a consequence of the zero in the upper-left corner of $\mm_d$.

We will now estimate the size of the $\Delta$ matrices appearing in Eq.(\ref{eq:thetadloop}) through a spurion analysis, as follows. The lagrangian will remain invariant under $SU(3)$ flavor transformations on the various quark fields, of the form
\begin{equation}
	d \rightarrow \crot_Q d , \qquad d' \rightarrow \crot_{Q'} d' , \qquad {\rm and} \qquad D \rightarrow \crot_D D , \qquad D' \rightarrow \crot_{D'} D' ,
\end{equation}
provided the various Yukawa couplings, as well as the vector-like mass matrix, similarly transform in an appropriate manner. The correct transformation rules for these objects are given by
\begin{equation} \begin{aligned}
	&y^T_d \rightarrow \crot_D^* y^T_d \crot_Q^\dagger \ ,	\qquad \qquad \qquad	& & y'^*_d \rightarrow \crot_{Q'}^* y'^*_d \crot_{D'}^\dagger \ , \\
	&\cm_d \rightarrow \crot_D^* \cm_d \crot_{D'}^\dagger \ ,	\qquad \qquad \qquad	& & \bar y_d \rightarrow \crot_D^* \bar y_d \crot_{D'}^\dagger \ .
\end{aligned}\end{equation}
On the other hand, the $\Delta$ matrices of Eq.(\ref{eq:thetadloop}) must similarly transform as follows:
\begin{equation}
	\Delta_{d' d} \rightarrow \crot_{Q'}^* \Delta_{d' d} \crot_Q^\dagger \ , \quad
	\Delta_{d' D'} \rightarrow \crot_{Q'}^* \Delta_{d' D'} \crot_{D'}^\dagger \ , \quad {\rm and} \quad
	\Delta_{D d} \rightarrow \crot_D^* \Delta_{D d} \crot_Q^\dagger \ .
\label{eq:Deltatrans}
\end{equation}

It is now straightforward to identify the leading objects that transform as in Eq.(\ref{eq:Deltatrans}) and contain a single insertion of $\bar y_d$. These are of the form
\begin{equation}
	\Delta_{d' d} \sim \frac{v v' v_\phi}{16 \pi^2} \left( y'^*_d \cm^{-1}_d \bar y_d \cm^{-1}_d y^T_d \right) ,
\end{equation}
whereas
\begin{equation}
	\Delta_{d' D'} \sim \frac{v' v_\phi}{16 \pi^2} \left( y'^*_d \cm^{-1}_d \bar y_d \right) , \qquad {\rm and} \qquad
	\Delta_{D d} \sim \frac{v v_\phi}{16 \pi^2} \left( \bar y_d \cm^{-1}_d y^T_d \right) .
\end{equation}
We have also included numerical factors to account for the loop suppression, as well as to take into account that the contribution to $\bar \theta$ must not diverge in the limits where either $v$ or $v'$ vanish. Plugging this back into Eq.(\ref{eq:thetadloop}), we find that all three terms give a contribution of the same size. Parametrically:
\begin{equation}
	\theta_d \sim \frac{v_\phi}{16 \pi^2} \Tr \left( \cm^{-1}_d \bar y_d \right) \sim \frac{ | \bar y_d | v_\phi}{16 \pi^2 M} .
\end{equation}
This result is consistent with the more detailed calculation of the contribution to $\bar \theta$ from the quark sector presented in appendix \ref{sec:apptheta}.

Requiring that $\bar \theta \lesssim 10^{-10}$ sets an upper bound on the typical size of the entries of the $\bar y$ matrices in the quark sector, of the form 
\begin{equation}
	\bar y \lesssim 10^{-8} \frac{M}{v_\phi} \lesssim 10^{-6} ,
\label{eq:maxbary}
\end{equation}
where in the last step we have assumed that $v_\phi \sim v'$, and have taken into account the upper bound on the see-saw scale $M$ as given in Eq.(\ref{eq:Mmax}).

This result bring us to the following conclusion: if the spontaneous breaking of parity also implies breaking $CP$, then any interaction between the quark and symmetry breaking sectors must be extremely weak. Fortunately, if $\bar y=0$ at tree-level, a non-zero value of $\bar y$ will not be generated radiatively. Indeed, $\bar y$ and the vector-like mass matrices $\cm$ are the only two parameters that violate the $\mathbb{Z}_2$ symmetry acting on the matter fields of the mirror sector. The breaking through $\cm$, however, is soft, and therefore will not translate into a non-zero $\bar y$ at the loop order. In this sense, a vanishing $\bar y$ is technically natural. 

\subsection{Spontaneously broken parity alone}

A less minimal possibility is to spontaneously break parity while preserving $CP$ through the addition of two scalar fields, $\sigma$ and $\sigma'$, whose vevs differ. This can be achieved if this symmetry breaking sector has a scalar potential of the form
\begin{equation}
	V_\sigma = - \frac{m_\sigma^2}{2} (\sigma^2 + \sigma'^2) + \frac{\lambda_1}{4} (\sigma^2 + \sigma'^2)^2 + \frac{\lambda_2}{4} \sigma^2 \sigma'^2 \ .
\end{equation}
where for simplicity we have forbidden cubic terms by imposing an additional $\mathbb{Z}_2$ symmetry. If $\lambda_2 > 0$, the vacua lie at $\langle \sigma \rangle = 0,$ $\langle \sigma' \rangle = \pm \sqrt{m_\sigma^2 / \lambda_1}$ and viceversa. This option is not viable for the Higgs potential itself, which requires both $v$ and $v'$ to be nonzero, but is perfectly adequate for an additional scalar sector.

Parity breaking can then be translated into the Higgs sector by writing appropriate couplings of the form
\begin{equation} \label{eq:higgsportal}
	V \supset \lambda_\sigma \left( \sigma^2 |H|^2 + \sigma'^2 |H'|^2 \right) + \lambda'_\sigma ( \sigma'^2 |H|^2 + \sigma^2 |H'|^2) \ .
\end{equation}
These terms are compatible with the generalized parity introduced in section \ref{sec:parity}, acting additionally as $\sigma \leftrightarrow \sigma'$. Provided $\lambda_\sigma \neq \lambda'_\sigma$, this will generate the soft term in Eq.(\ref{eq:scalarsoft}) proportional to $\lambda_\sigma - \lambda'_\sigma$. For example, in the vacuum with $\langle \sigma' \rangle \neq 0$, $v' \gg v$ corresponds to
\begin{equation}
(\lambda'_\sigma - \lambda_\sigma) \langle \sigma' \rangle^2 \sim \kappa v'^2 .
\end{equation}
As this scenario breaks $P$ without breaking $CP$ (and the additional $\mathbb{Z}_2$ symmetry acting on the $\sigma$s forbids marginal couplings between $\sigma, \sigma'$ and fermion bilinears), there are no significant additional contributions to the neutron EDM. There is, of course, the possibility of collider signatures coming from the Higgs portal coupling in Eq.(\ref{eq:higgsportal}), most notably mixing between the Higgs and the scalar that acquires a vev, as well as invisible decays of the Higgs if kinematically allowed. The two scalars acquire masses of order $\sqrt{2 \lambda_1} \langle \sigma' \rangle \sim \sqrt{\lambda_1} v'$ and $\sqrt{\lambda_2/2} \langle \sigma' \rangle \sim \sqrt{\lambda_2} v'$, respectively, and it is certainly possible for one to be lighter than half the Higgs mass depending on the values of $\lambda_{1,2}$.

\section{Strong CP and quantum gravity}
\label{sec:HDO}

As we discussed in the Introduction, the strong CP problem arises out of the difficulty of reconciling the smallness of $\bar \theta$ with the $\co(1)$ violation of both parity and $CP$ by the electroweak sector. In turn, all attempts to address this puzzle are themselves based on the introduction of an additional symmetry beyond those of the SM. However, there is strong evidence that within a theory of quantum gravity, global symmetries cannot be exact --- they must be either broken, or gauged. The origin of this statement goes back a long way \cite{Zeldovich:1976vq,Zeldovich:1977be,Banks:1988yz,Giddings:1987cg,Lee:1988ge,Abbott:1989jw,Coleman:1989zu,Kallosh:1995hi,Banks:2010zn}, and to some extent it has recently been established \cite{Harlow:2018tng,Harlow:2018jwu}. Of course, the single most pressing issue for phenomenology is to establish a lower bound on the amount of global symmetry violation that must be present in the IR. Attempts at finding such a ``universal'' lower bound have been made \cite{Fichet:2019ugl,Daus:2020vtf}, but a fully satisfactory answer remains elusive. Absent a full understanding of how quantum gravity affects global symmetries at low energies, we can at least attempt to assess the robustness of an EFT against global symmetry violation by considering the impact of HDOs suppressed by the appropriate power of $M_{Pl}$.  This both constrains the viable parameter space of parity solutions to strong CP and illustrates the sense in which $P$, rather than $U(1)_{PQ}$, provides a solution to the strong CP problem that is robust against the expected intrusion of quantum gravity. Beyond imposing constraints, these higher-dimensional operators also lead to new experimental signatures associated with the spontaneous breaking of parity, which we explore in section \ref{sec:cosmology}.

\subsection{Constraints from Planck-suppressed operators}

The observation that the breaking of global symmetries by quantum gravity can have a profound impact on the validity of the QCD axion solution to strong CP was first made in \cite{Barr:1992qq,Kamionkowski:1992mf,Holman:1992us,Ghigna:1992iv}. Planck-suppressed HDOs that violate the $U(1)_{PQ}$ symmetry carried by the field $\Phi$, the phase of which is the axion, are of the form
\begin{equation}
	\cl \supset \frac{\eta}{M_{Pl}^{d-4}} |\Phi|^{d-n} \Phi^n + {\rm h.c.}
\end{equation}
Here, $d$ is the operator dimension, $n$ its units of $U(1)_{PQ}$ charge (so $n \geq 1$ in order to break the symmetry), and $\eta$ a coupling that will in general feature arbitrary real and imaginary parts. HDOs of this form contribute to the axion potential, and, in general, will displace the axion vev away from the value leading to a small $\bar \theta$. Following \cite{Kamionkowski:1992mf}, requiring that the shift in the axion vev is small enough so as not to spoil the solution to strong CP translates into the following upper bound
\begin{equation}
	|\eta| \left( \frac{f_a}{\sqrt{2} M_{Pl}} \right)^d \lesssim 10^{-81} \bar \theta \lesssim 10^{-91} ,
\end{equation}
where $f_a$ is the scale of $U(1)_{PQ}$ spontaneous symmetry breaking (alternatively, the axion decay constant), which is experimentally constrained to be between $10^8$ and $10^{17}$ GeV \cite{Zyla:2020zbs}. Focusing on operators of dimension $d=5$, this translates into an upper bound on the size of $\eta$, of the form
\begin{equation}
	|\eta| \lesssim 10^{-55} \left( \frac{10^{12} {\rm GeV}}{f_a} \right)^5 \left( \frac{\bar \theta}{10^{-10}} \right) \ .
\end{equation}
In other words, for all experimentally allowed values of the axion decay constant, the $U(1)_{PQ}$ symmetry must remain an approximate global symmetry to an exceptional degree.
This is clearly one of the most significant drawbacks of the axion solution to strong CP.

In the remainder of this section we study the effect of Planck-suppressed HDOs on parity solutions to the strong CP problem. We consider separately the cases where parity is global or gauged. The nature of the HDOs under consideration will be different, but in both cases we will see that even $\co(1)$ coefficients are compatible with solving strong CP.

\subsubsection*{Parity as a global symmetry}

If we regard parity as a global symmetry, then we must consider the effect of HDOs that explicitly violate $P$. The relevant dimension-5 HDOs were already identified in \cite{Berezhiani:1992pq}, and they are of the form
\begin{equation}
    \cl \supset \frac{1}{M_{Pl}} \left[ (\alpha_u)_{ij} (H' Q_i')(H Q_j) + (\alpha_d)_{ij} (H'^\dagger Q_i')(H^\dagger Q_j) \right] + \textrm{h.c.}
\label{eq:HDOglobal}
\end{equation}
Notice that if $\alpha_f = \alpha_f^\dagger$ then the above terms would be parity-symmetric. In general, however, the $\alpha_f$'s will \emph{not} be hermitian, and it is under this assumption that we proceed.

Setting the Higgs to their vevs, Eq.(\ref{eq:HDOglobal}) leads to a correction to the quark mass matrix that, for arbitrary $\alpha_f$'s, does not respect generalized parity. The leading contribution to $\bar \theta$ will come from the contributions to the up- and down-quark masses, which are of the form
\begin{equation}
	\delta m_u \simeq \frac{vv' (\alpha_u)_{11}}{2 M_{Pl}}, \qquad {\rm and} \qquad \delta m_d \simeq \frac{vv' (\alpha_d)_{11}}{2 M_{Pl}} .
\end{equation}
In turn,
\begin{equation}
	\theta_q \simeq \frac{\Imag (\delta m_u)}{m_u} + \frac{\Imag (\delta m_d)}{m_d} \sim 10^5 \frac{|\alpha| v'}{2 M_{Pl}} ,
\end{equation}
where in the last step we have used $m_u /v \sim m_d / v \sim 10^{-5}$. Requiring that the above contribution is smaller than the current bound on $\bar \theta$ translates into an upper bound on the parity breaking scale:
\begin{equation}
	v' \lesssim \frac{20 \ {\rm TeV}}{|\alpha|} \left( \frac{\bar \theta}{10^{-10}} \right) .
\end{equation}
Notice this upper bound is (just) compatible with the lower bound $v' \gtrsim 18$ TeV from direct searches of $W'$ gauge bosons, as discussed in section \ref{sec:irreducible}. As a result, if \emph{global} generalized parity is responsible for solving strong CP, an $\co(1)$ violation of the symmetry due to gravitational effects would imply a contribution to $\bar \theta$ accessible in near-future experiments.

\subsubsection*{Parity as a gauge symmetry}

If parity is instead a gauge symmetry of the underlying theory, HDOs that explicitly violate $P$ are therefore not allowed. Planck-suppressed operators such as those in Eq.(\ref{eq:HDOglobal}) might still be generated, but only with $\alpha_f = \alpha_f^\dagger$, and therefore will not contribute to $\bar \theta$. Instead, the operators of interest must be proportional to the source of spontaneous symmetry breaking. If the latter takes place via the vev of a pseudo-scalar, as discussed in section~\ref{sec:PandCP}, then there are two dimension-5 HDOs that satisfy this requirement, namely:
\begin{equation}
	\cl \supset \eta_s \frac{\phi \alpha_s}{4 \pi M_{Pl}} \Tr \left( G^a \tilde G^a \right) ,
\label{eq:HDOthetas}
\end{equation}
and
\begin{equation}
	\cl \supset \frac{i \phi}{M_{Pl}} \left\{ (\zeta_u)_{ij} Q_i H U_j + (\zeta'_u)_{ij} Q'_i H' U'_j + (\zeta_d)_{ij} H^\dagger Q_i D_j + (\zeta'_d)_{ij} H'^\dagger Q'_i U'_j \right\} + {\rm h.c.} ,
\label{eq:HDOthetaq}
\end{equation}
with $\eta_s \in \mathbb{R}$, and $\zeta'_f = \zeta_f^*$ so as to satisfy generalized parity.

The operator of Eq.(\ref{eq:HDOthetas}) will generate a contribution to $\theta_s$ after spontaneous symmetry breaking, of the form
\begin{equation}
	\theta_s \simeq \frac{\eta_s v_\phi}{M_{Pl}}.
\end{equation}
Assuming that $v_\phi \sim v'$, demanding that this contribution is smaller than the current bound on $\bar \theta$ leads to an upper bound on the parity breaking scale
\begin{equation}
	v_\phi \sim v' \lesssim \frac{10^9 \ {\rm GeV}}{\eta_s} \left( \frac{\bar \theta}{10^{-10}} \right) ,
\end{equation}
which is clearly well above current bounds on $v'$.

At the same time, once $\phi$ gets its vev, the operator of Eq.(\ref{eq:HDOthetaq}) leads to an extra contribution to the Yukawa couplings of the up- and down-type quarks in the SM and mirror sectors that are \emph{not} parity-symmetric. In turn, this will lead to an additional contribution to the mass eigenvalues of the light quarks which will in general contain an imaginary component. For example, in the down quark sector
\begin{equation}
	\Imag (\delta m_{d_i}) \simeq v v' \Imag \left\{ \frac{i v_\phi}{M_{Pl}} \sum_j \frac{(y_d)^*_{ij} (\zeta_{d})_{ij}}{m_{D_j}} \right\} \sim \frac{v v'}{M} \frac{|\zeta_d| v_\phi}{M_{Pl}} |(y_d)_{i \star}| ,
\end{equation}
where $|(y_d)_{i \star}|$ refers to the typical size of the entries in the $i$-th row of the $y_d$ matrix. The leading contribution to $\theta_q$ will come from the up and down quarks.
In total:
\begin{equation}
	\theta_q \simeq \frac{\Imag (\delta m_u)}{m_u} + \frac{\Imag (\delta m_d)}{m_d} \simeq \frac{v_\phi}{M_{Pl}} \frac{v v'}{M} \left\{ \frac{|\zeta_u| (y_u)_{1 \star}}{m_u} + \frac{|\zeta_d| (y_d)_{1 \star}}{m_d} \right\} .
\end{equation}
Taking into account the upper bound on the entries of the Yukawa couplings necessary to reproduce the light quark masses (see Eq.(\ref{eq:ymax})), as well as the requirement that $v' \lesssim M$ in order to implement the see-saw mechanism, the previous equation implies
\begin{equation}
	\theta_q \lesssim 10^2 \frac{ |\zeta| v_\phi}{M_{Pl}} \ ,
\end{equation}
where we have set $m_u/v \sim m_d/v \sim 10^{-5}$, and have assumed that $|\zeta_u| \sim |\zeta_d|$. In turn, taking $v_\phi \sim v'$, this sets an upper bound on the scale of spontaneous symmetry breaking:
\begin{equation}
	v_\phi \sim v' \lesssim \frac{10^7 \ {\rm GeV}}{|\zeta|} \left( \frac{\bar \theta}{10^{-10}} \right) .
\label{eq:v'max_Pgauged}
\end{equation}
As before, this is fully compatible with current experimental bounds on the parity-breaking scale, even for $\co(1)$ coefficients of the corresponding HDOs.

\subsection{Gravitational waves from the spontaneous breaking of parity}
\label{sec:cosmology}

Beyond providing additional constraints on the parameter space of parity solutions to strong CP, the expected effects of gravity also introduce new experimental signatures. Here we highlight one possibility, namely the impact of HDOs when parity is a spontaneously broken global symmetry. The spontaneous breaking of discrete symmetries can lead to the formation of a network of domain walls in the early universe, provided the reheating temperature after inflation is above the scale of spontaneous symmetry breaking \cite{Kibble:1976sj}. If the spontaneously broken symmetry is global, but otherwise exact, a domain wall configuration interpolates between two distinct vacua that are degenerate, making these defects topologically stable objects. The formation of such networks can be fatal on two grounds. On the one hand, the energy density in domain walls redshifts slower than that of matter or radiation, and would eventually dominate the universe's energy budget. If this happened before the current epoch, the rapid expansion of the subsequent domain-wall-dominated era would be at odds with observation. On the other hand, even if only a subdominant component of the total energy density was in the form of domain walls today, their effect on large-scale density fluctuations rules out defects with characteristic scales above $\sim 1 \ {\rm MeV}$ \cite{Zeldovich:1974uw}. These considerations are often referred to as the ``domain wall problem" of theories with spontaneously broken discrete symmetries.

These problems are largely solved when we take into consideration that, within a theory of quantum gravity, we expect all symmetries to be either broken or gauged \cite{Zeldovich:1976vq,Zeldovich:1977be,Banks:1988yz,Giddings:1987cg,Lee:1988ge,Abbott:1989jw,Coleman:1989zu,Kallosh:1995hi,Banks:2010zn,Harlow:2018tng,Harlow:2018jwu} --- an expectation that includes spacetime symmetries \cite{Harlow:2018tng,Harlow:2018jwu}. In this context, the domain wall network is unstable, rendering its earlier formation largely unproblematic. Moreover, the significant amount of gravitational radiation emitted in the process results in a stochastic gravitational wave background that may be within reach of current and future observatories. We discuss this possibility in the remainder of this section.

We will focus first on the scenario where parity is a global symmetry that is only explicitly broken by gravitational effects. At low energies, the symmetry-breaking dynamics will enter the effective potential for $\phi$ through HDOs that violate parity. One such operator is of the form
\begin{equation}
	V \supset \epsilon \frac{\phi^5}{M_{Pl}} \ .
\label{eq:HDOphi}
\end{equation}
This breaks the degeneracy between the two previously degenerate vacua, corresponding to $\langle \phi \rangle = \pm v_\phi$. Parametrically, the energy difference now reads
\begin{equation}
	\delta V  \sim \frac{\epsilon v_\phi^5}{M_{Pl}} \ .
\end{equation}

\begin{figure}
  \centering
  \includegraphics[scale=0.8]{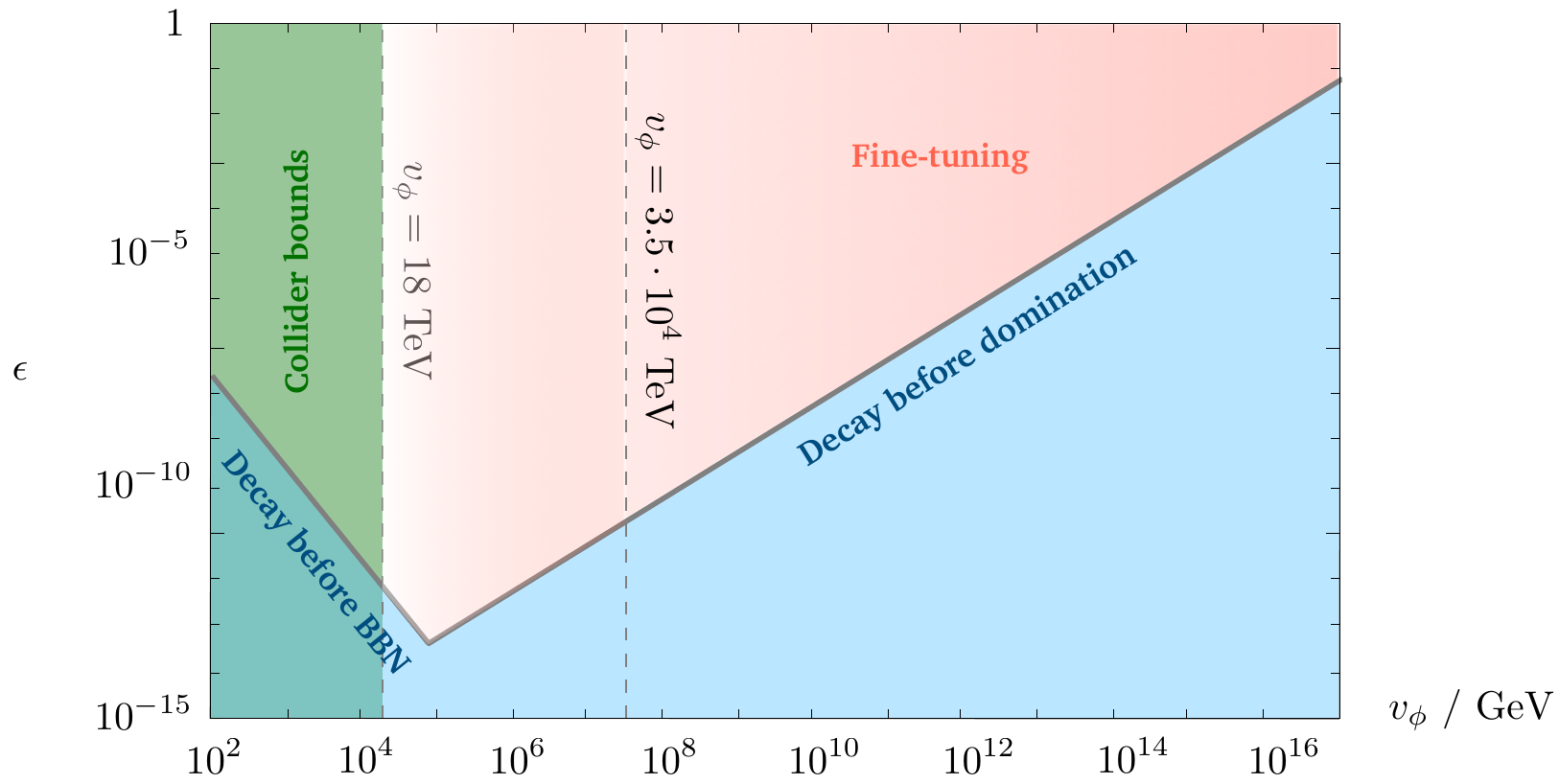}
  \caption{Constraints on the size of the coefficient of the Planck-suppressed HDO of Eq.(\ref{eq:HDOphi}), as a function of the scale of spontaneous symmetry breaking, $v_\phi$. The region $v_\phi \lesssim 18$ TeV (green) is in conflict with direct bounds on the mass of $W'$ and $Z'$ resonances, under the assumption that $v_\phi \sim v'$, as discussed in section \ref{sec:irreducible}. Values of $\epsilon$ that are too small (blue) do not destabilize the domain wall network early enough to either avoid a domain-wall-dominated era, or to ensure collapse before the onset of BBN, and are therefore ruled out. The region of parameter space in pink is experimentally allowed, but the level of fine-tuning in the electroweak sector worsens as $v_\phi$ is increased (corresponding to a darker shade). The dashed line at $v_\phi \simeq 3.5 \cdot 10^4$ TeV corresponds to a fine-tuning of $\co(10^{-10})$ in the electroweak sector. (For illustration, we have set the quartic coupling in the pseudo-scalar potential to be $\kappa_\phi = 1$ in this plot.)}
  \label{fig:epsilonmin}
\end{figure}

If the reheating temperature is above the scale of spontaneous symmetry breaking, then we expect that a network of domain walls will be formed once the temperature of the universe drops bellow $T \sim v_\phi$ \cite{Kibble:1976sj}. Numerical \cite{Press:1989yh,Garagounis:2002kt,Oliveira:2004he,Avelino:2005kn,Leite:2011sc} and analytical \cite{Hindmarsh:1996xv,Hindmarsh:2002bq} studies suggest that, shortly after formation, the network evolves according to a scaling solution, with $\rho_{\rm DW} (t) \simeq \sigma/t$, and typical domain wall size comparable to the Hubble scale $H(t)^{-1}$. $\sigma$ corresponds to the tension of the domain walls, which in our model is of the form $\sigma \sim \sqrt{\kappa_\phi} v_\phi^3$, where $\kappa_\phi$ refers to the quartic coupling in the $\phi$ potential. Two competing effects determine the network's subsequent evolution. On the one hand, the pressure difference between the two vacua exerts a force per unit area of order $\sim \delta V$. On the other, the tension per unit area acting on a wall with curvature radius $R$ is $\sim \sigma / R$. In the scaling regime, $R (t) \sim H(t)^{-1} \sim t$ (assuming the universe is radiation dominated), and therefore the effect of tension decreases with time. Eventually, the pressure difference between the two vacua dominates, causing the network to collapse at a time of order
\begin{equation}
	t_{\rm coll.} \sim \frac{\sigma}{\delta V} \sim \frac{\sqrt{\kappa_\phi} M_{Pl}}{\epsilon v_\phi^2} \ .
\end{equation}
Clearly, the domain wall network could be very long-lived if $\epsilon \lll 1$. The requirement that collapse takes place either before the universe becomes domain wall dominated, or before the start of BBN (so as to avoid energy injection into the SM plasma that would disrupt light element formation), sets a lower bound on $\epsilon$ as a function of the spontaneous symmetry breaking scale. This is depicted in figure~\ref{fig:epsilonmin}, where the BBN and domain-wall-domination restrictions dominate for values of $v_\phi$ below and above $\sim 73$ TeV respectively.
As can be appreciated in the figure, in the region of parameter space where the fine-tuning is better than $10^{-10}$ (that is, $v_\phi \sim v' \lesssim 3.5 \cdot 10^{4}$ TeV), $\epsilon$ may be as small as $\co(10^{-13})$.

The collapse of a domain wall network leads to the production of gravitational waves \cite{Vilenkin:1981zs,Vachaspati:1984yi}. On dimensional grounds, one would expect the energy density in gravitational radiation to be of the form $\rho_{\rm gw} \sim G_N \sigma^2$ (the mandatory power of $G_N$ times the necessary factors of $\sigma$ to make up dimensions), an expectation that is largely upheld by numerical analysis \cite{Gleiser:1998na,Dufaux:2007pt,Hiramatsu:2010yz,Kawasaki:2011vv}. The resulting gravitational wave spectrum has an extended shape, peaking at a frequency corresponding to the Hubble size at the time of collapse (corresponding to the typical size of the domain walls), and falling off as $1/f$ for larger frequencies.
At the present epoch, the peak frequency of the gravitational wave signal is given by
\begin{equation} \begin{aligned}
f_* 	& \simeq 10^{-9} \ {\rm Hz} \left( \frac{T_{\rm coll.}}{10^{-2} \ {\rm GeV}} \right) \left( \frac{g_* (T_{\rm coll.})}{10} \right)^{1/6} \\
	& \sim 10^{-9} \ {\rm Hz} \left( \frac{v_\phi}{18 \ {\rm TeV}} \right)	\left( \frac{\epsilon}{10^{-12}} \right)^{1/2}	\left( \frac{1}{\kappa_\phi} \right)^{1/4} ,
\label{eq:fpeak}
\end{aligned} \end{equation}
and the energy density in gravitational radiation at frequency peak reads \cite{Saikawa:2017hiv}
\begin{equation} \begin{aligned}
	\Omega_{\rm gw} h^2 (f_*) & \simeq 2 \cdot 10^{-10} \left( \frac{\sigma}{(20 \ {\rm TeV})^3} \right)^2 \left( \frac{10^{-2} \ {\rm GeV}}{T_{\rm coll.}} \right)^4 \left( \frac{10}{g_*(T_{\rm coll.})} \right)^{4/3} \\
						& \sim 10^{-10} \left( \frac{v_\phi}{18 \ {\rm TeV}} \right)^2 \left( \frac{10^{-12}}{\epsilon} \right)^2 \left( \frac{\kappa_\phi}{1} \right)^2 ,
\label{eq:Omegapeak}
\end{aligned} \end{equation}
where $T_{\rm coll.}$ refers to the temperature of the SM plasma at a time $t_{\rm coll.}$, and we have assumed that network collapse takes place during radiation domination.\footnote{ In the second steps of Eq.(\ref{eq:fpeak}) and (\ref{eq:Omegapeak}), we have substituted $T_{\rm coll.}$~by the corresponding expression in terms of the model's fundamental parameters, while ignoring a weak dependence on $g_*$.}

\begin{figure}
  \centering
  \includegraphics[scale=0.8]{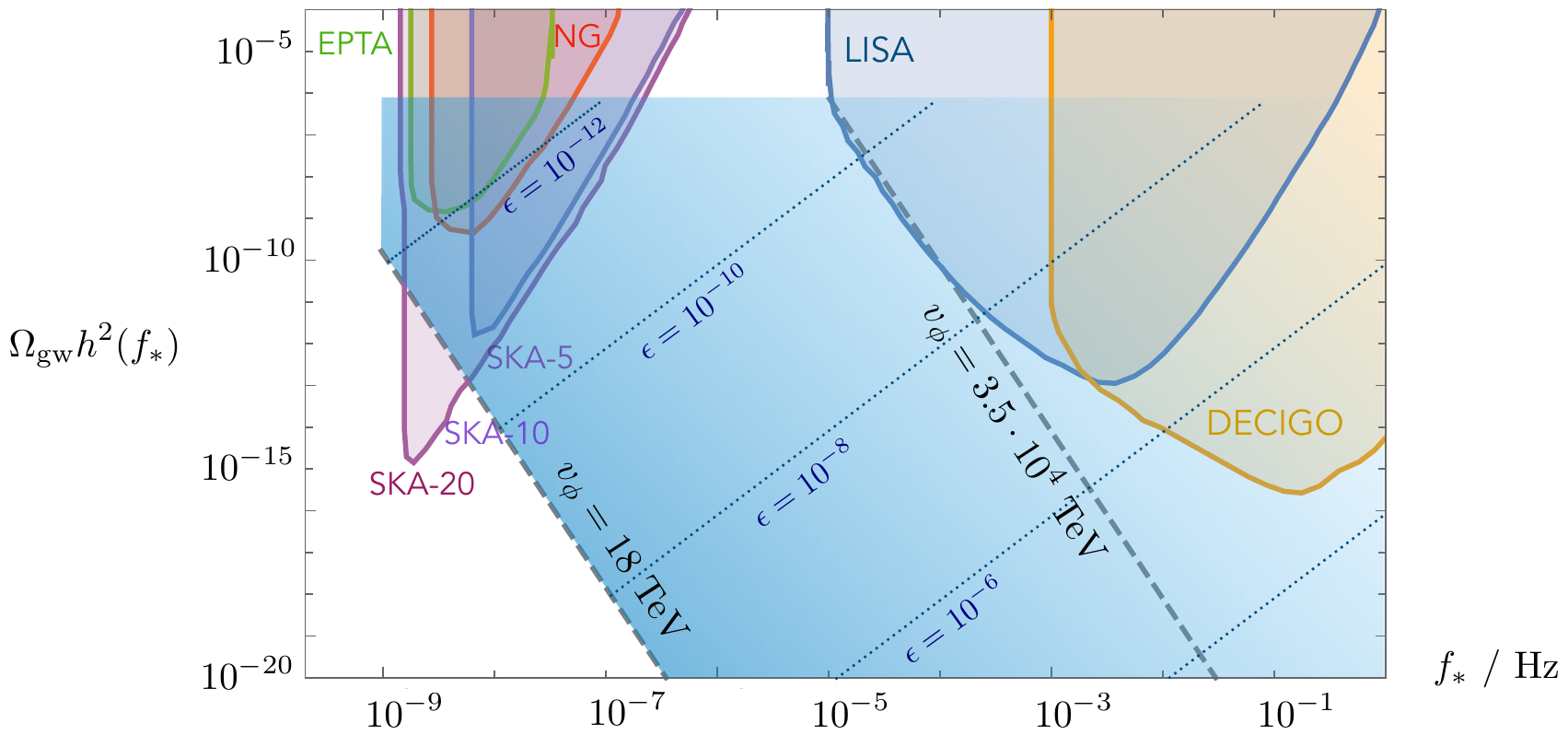}
  \caption{In blue, region of parameter space where the frequency peak of the stochastic gravitational wave signal, $f_*$, as well as the corresponding energy density, $\Omega_{\rm gw} h^2 (f_*)$, can fall in light of the experimental constraints on the various model parameters summarized in figure \ref{fig:epsilonmin}. Dotted lines correspond to constant $\epsilon$. The region to the right of the dashed line corresponding to $v_\phi = 3.5 \cdot 10^4$ TeV features a level of fine-tuning worse than $1$ part in $10^{10}$, and it is therefore less attractive. Sensitivity curves for a variety of gravitational wave experiments are shown, including the pulsar timing arrays EPTA \cite{Lentati:2015qwp}, NANOGrav \cite{Arzoumanian:2018saf}, SKA \cite{Janssen:2014dka} (observation time of 5, 10 and 20 years as indicated), as well as the space-based interferometers LISA \cite{Audley:2017drz}, and DECIGO \cite{Seto:2001qf}. (For illustration, we have set the quartic coupling in the pseudo-scalar potential to be $\kappa_\phi = 1$ in this plot.)}
  \label{fig:gwsignal}
\end{figure}

Figure~\ref{fig:gwsignal} shows the region that can be spanned by the peak of the stochastic gravitational wave background in the $f_*$ vs.~$\Omega_{\rm gw} h^2 (f_*)$ plane, together with the sensitivity curves of a number of gravitational wave experiments. The lower bound on $\epsilon$ depicted in figure \ref{fig:epsilonmin} translates into a lower bound on $f_*$ for each value of the symmetry breaking scale (e.g.~$f_* \gtrsim 10^{-9}$ Hz for $v_\phi \simeq 18$ TeV). As can be seen in figure \ref{fig:gwsignal}, a region of parameter space with low $v'$ falls within reach of gravitational wave observatories probing the low frequency regime such as SKA \cite{Janssen:2014dka}, NANOGrav \cite{Arzoumanian:2018saf}, and the EPTA \cite{Lentati:2015qwp}.
As $\epsilon$ is increased, the collapse of the domain wall network occurs earlier, further suppressing the current value of the energy density in gravitational radiation by the corresponding redshift factor.

Our discussion so far applies in the context of global discrete symmetries provided that they either do not descend from a continuous symmetry, or that, if they do, the symmetry breaking scale of the continuous factor is above the reheating temperature, so that a network of cosmic strings is not formed in the early universe. On the other hand, if the reheating temperature is larger than the scale set by the tension of the strings, $\mu$, then a string network will be formed first, with the strings later joined by domain walls. The entire string-wall network now evolves together, and the problem features an additional time scale, given by
\begin{equation}
	t_* \sim \frac{\mu}{\sigma} \ .
\end{equation}
At $t \sim t_*$, the force per unit length on a string of radius $R(t) \sim t$, given by $\sim \mu / t$, becomes comparable to the wall tension. The system then becomes dominated by the tension of the domain walls, causing the network to shrink, and break down into pieces that will further decay into gravitational waves (or, potentially, also massive particles, depending on their size and the relevant particle spectrum) \cite{Vilenkin:1981zs,Vachaspati:1984yi}. If this timescale is shorter than $t_{\rm coll.}$, the earlier destruction of the network of defects could move any potential gravitational wave signal into an unobservable regime.

The discussion of the previous paragraph is especially relevant if parity is instead realized as a gauge symmetry, for which explicit breaking is no longer allowed. Na\"ively, one would hope that the gauge case would be cosmologically more benign: the gauge equivalence of the two vacua makes them no longer distinct, eliminating the topological stability of the domain walls. Indeed, domain walls can be destroyed by a process in which a string loop is nucleated on the wall, further growing to destroy the entire defect. However, the corresponding nucleation probability is proportional to $e^{-\mu^3 / \sigma^2}$ \cite{Preskill:1992ck}, which will be exceedingly small for any reasonable separation of scales between the string and wall tensions, therefore rendering gauge domain walls effectively stable. It is therefore crucial that the reheating temperature is above the string tension scale, so that a string network is formed that can later result in the entire collapse of the subsequent string-wall network. The cosmological implications, as well as potential gravitational wave signatures, of a discrete parity symmetry that is gauged will be further explored in future work.

\section{Conclusions}
\label{sec:conclusions}

The strong CP problem remains one of the great naturalness problems of the Standard Model, and is perhaps the most compelling in light of its resistance to straightforward anthropic explanations. Fully satisfying solutions to the problem remain elusive given the expected violation of global symmetries in a theory of quantum gravity, which demands extensive effort to protect the Peccei-Quinn symmetry underlying axion-based approaches. In this work we have pursued a possibility that is more transparently robust against the effects of quantum gravity, revisiting parity-based solutions to the strong CP problem. Our approach highlights the experimental signatures associated with the most natural regions of parameter space in these models, as well as ancillary signatures that are dependent upon the detailed mechanism of parity breaking. 

The notion of naturalness within this parameter space is governed by the tuning associated with the separation of scales of $SU(2)_L$ and $SU(2)_R$ breaking, which are related by generalized parity. Given this tuning, ``see-saw'' vector-like masses for the $SU(2)$-singlet fermions play a key role in allowing the scale of $SU(2)_R$ breaking to be lowered toward its most natural value consistent with experimental constraints. Within this framework, the LHC provides the strongest test of natural parity-based solutions to the strong CP problem, probing the scale of $SU(2)_R$ breaking through searches for $W'$ and $Z'$ vector bosons as well as vector-like quarks and additional Higgs bosons. This leaves parity solutions tuned at the $\sim 10^{-3}$ level, which while not fully natural remains a significant improvement in explaining the observed $\bar \theta \lesssim 10^{-10}$. The extended reach for heavy resonances at future colliders such as FCC-hh will decisively test these parity solutions at the level of $\sim 10^{-5}$ tuning. Constraints on new sources of flavor violation play a complementary role, with additional sensitivity to the scale of vector-like fermions and the underlying model of flavor. 

The detailed mechanism of parity breaking gives rise to additional signatures within reach of near-future tabletop experiments and gravitational wave observatories. Soft parity and $CP$-violating terms give rise to EDMs for elementary fermions at one-loop, both quarks and charged leptons, which provide a pathway to discovery in precision searches for $CP$-violation in molecular systems. Spontaneous violation of parity and $CP$ through the vev of a pseudo-scalar gives rise to additional one-loop contributions to $\bar \theta$, which provides an additional pathway to discovery and already requires the source of parity violation to be sequestered from the quark sector (albeit in a technically natural way). The expected violation of global symmetries in a theory of quantum gravity further shapes the viable parameter space and potential experimental signatures through the impact of various Planck-suppressed operators whose form depends on the underlying parity-breaking mechanism. If parity is a global symmetry that is broken both spontaneously (by a pseudo-scalar vev) and explicitly (by gravitational effects), collapse of the domain wall network associated with the spontaneous breaking of parity can generate a gravity wave signal accessible at low-frequency gravitational wave observatories. In this respect, the violation of global symmetries by gravitational effects is a feature of parity-based solutions to the strong CP problem, rather than a bug. Taken together, these experimental opportunities warrant further exploration of generalized parity as a solution to strong CP.

\section*{Acknowledgments}
We thank P.~Draper and A.~Jayich for useful conversations. The research of IGG is funded by the Gordon and Betty Moore Foundation through Grant GBMF7392, and in part by the National Science Foundation under Grant No.~NSF PHY-1748958. The research of NC, GK, and AM is supported in part by the Department of Energy under the grant DE-SC0011702 and the Cottrell Scholar Program through the Research Corporation for Science Advancement.


\appendix

\addtocontents{toc}{\protect\setcounter{tocdepth}{1}} 

\section{Mass eigenstates}
\label{sec:massbasis}

\subsection{Gauge and Higgs sectors}
\label{sec:appgauge}

With the gauge group of Eq.(\ref{eq:gaugegroup}), and the Higgs sector specified in table~\ref{tab:model}, spontaneous symmetry breaking takes place in two steps, as follows
\begin{equation}
	SU(2)_L \times SU(2)_R \times U(1)_{\hat Y} \xrightarrow{v' \neq 0} SU(2)_L \times U(1)_Y \xrightarrow{v \neq 0} U(1)_{EM} .
\end{equation}
The physical spectrum contains SM-like $Z$, $W^\pm$, and $\gamma$ gauge bosons, as well as exotic $Z'$ and $W'^\pm$ excitations. At tree-level, no mixing occurs in the charged gauge boson sector, and the mass eigenstates are given in terms of the gauge eigenbasis by the usual expression:
\begin{equation}
	W^\pm = \frac{1}{\sqrt{2}} (W^1 \mp i W^2) \ ,
\end{equation}
and similarly in the $W'$ sector. Tree-level masses are of the form $m_W = gv/2$ and $m_{W'} = gv'/2$, where we have assumed that $g'=g$, as mandated by generalized parity. By contrast, in the neutral gauge boson sector mixing between SM and mirror fields takes place already at tree-level. At zeroth order in a $v/v'$ expansion, the gauge eigenstates can be written in the mass eigenbasis as follows
\begin{equation}
	\begin{pmatrix} W'^3_\mu \\ W^3_\mu \\ \hat B_\mu \end{pmatrix}
	=
	\begin{pmatrix}	\frac{\sqrt{ \cos 2 \theta_w}}{\cos \theta_w} & & & - \sin \theta_w \tan \theta_w & & & \sin \theta_w \\
				0 & & & \cos \theta_w & & & \sin \theta_w \\
				- \tan \theta_w & & & - \tan \theta_w \sqrt{ \cos 2 \theta_w} & & & \sqrt{ \cos 2 \theta_w} \end{pmatrix}
	 \begin{pmatrix} Z'_\mu \\ Z_\mu \\ A_\mu \end{pmatrix},
\label{eq:rotationgauge}
\end{equation}
where $\sin^2 \theta_w \simeq 0.231$ as usual. Corrections to the above expression arise at $\co(v^2/v'^2)$. Masses for the SM-like $Z$ and mirror $Z'$ are given by
\begin{equation}
	m_Z = \frac{g v}{2 \cos \theta_w} + \co \left( \frac{v^2}{v'^2} \right), \qquad {\rm and} \qquad
	m_{Z'} = \frac{g v' \cos \theta_w}{2 \sqrt{\cos 2\theta_w}} + \co \left( \frac{v^2}{v'^2} \right) .
\end{equation}

After electroweak symmetry breaking, the Higgs sector consists of two real scalar fields, $h$ and $h'$, with masses given by $m_h \simeq 2 \sqrt{\kappa} v$ and $m_{h'} \simeq \sqrt{2 \lambda} v'$. Rotating from the gauge to the mass eigenbasis can be performed as follows
\begin{equation}
	\begin{pmatrix} h \\ h' \end{pmatrix} \rightarrow \begin{pmatrix} \cos \alpha & & \sin \alpha \\ - \sin \alpha & & \cos \alpha \end{pmatrix} \begin{pmatrix} h \\ h' \end{pmatrix} \ ,
\end{equation}
with mixing angle $\alpha \sim v/v'$.

\subsection{Fermion sector}
\label{sec:appfermions}

Rotating from the flavor to the mass eigenbasis in the fermion sector requires solving the eigenvalue problem for the $6 \times 6$ matrices $\mathbb{M}_f^\dagger \mathbb{M}_f$, and $\mathbb{M}_f \mathbb{M}_f^\dagger$, with $\mathbb{M}_f$ as given in Eq.(\ref{eq:M6x6}). This can be conveniently done as a perturbation expansion in $v/M, v'/M \ll 1$. In this section, we summarize the relevant results of this procedure. We focus first on the down-quark and lepton sectors (although we will use notation appropriate to the down-quark sector, we emphasize that identical results apply for leptons). The singularities of the up sector as related to the top quark merit a separate discussion that we present later.

\subsubsection*{Down-type quarks and leptons} 

The mass eigenvalues in the down-quark sector can be found by diagonalizing the two $3 \times 3$ matrices
\begin{equation}
	\frac{v'v}{2} y'^*_d \cm^{-1}_d y^T_d , \qquad {\rm and} \qquad \cm_d .
\label{eq:dquarksmasses}
\end{equation}
In full generality, i.e.~without yet imposing generalized parity, the above matrices are not necessarily hermitian, and two unitary matrices are needed in order to bring them into real diagonal form. This corresponds to the unitary transformations
\begin{equation}
	d \rightarrow \co^\dagger_d d , \ d' \rightarrow \co^\dagger_{d'} d' , \qquad {\rm and} \qquad D \rightarrow \co^\dagger_D D , \ D' \rightarrow \co^\dagger_{D'} D' .
\label{eq:rotationsd}
\end{equation}
By definition, the rotation matrices are such that
\begin{equation}
	\mathbb{m}_d \equiv \co^*_{d'} \left( \frac{v'v}{2} y'^*_d \cm_d^{-1} y_d^T \right) \co^\dagger_d = {\rm diag} (m_{d_i}) \ ,
\label{eq:deffmd}
\end{equation}
and
\begin{equation}
	\mathbb{m}_D \equiv \co^*_D \cm_d \co^\dagger_{D'} = {\rm diag} (m_{D_i} ) \ ,
\end{equation}
where $m_{d_i}$ and $m_{D_i}$ are the masses of the SM and exotic heavy quarks respectively. As advertised in section~\ref{sec:fermionmasses}, we will make the simplifying assumption that all three mirror quarks appear at a common scale $m_{D_i} \sim M \gg v, v'$.
Imposing generalized parity makes both matrices in Eq.(\ref{eq:dquarksmasses}) hermitian. In this case, a single unitary matrix suffices to make them diagonal, and we have $\co_{d'} = \co^*_d$ and $\co_D = \co^*_{D'}$.

It is convenient to define two new matrices corresponding to the Yukawa couplings in this new basis
\begin{equation}
	\tilde y_d \equiv \co^*_d y_d \co^\dagger_D , \qquad {\rm and} \qquad \tilde y'_d \equiv \co_{d'} y'_d \co^T_{D'} \ .
\label{eq:deftildey}
\end{equation}
With this definition, the tree-level masses of the SM-like fermions read
\begin{equation}
	m_{d_i} = \frac{v v'}{2} \sum_j \frac{(\tilde y'^*_d)_{ij} (\tilde y_d)_{ij} }{ m_{D_j} } = \frac{v v'}{2} \sum_j \frac{ | (\tilde y_d)_{ij} |^2 }{ m_{D_j} } \ ,
\end{equation}
where the last step holds provided we impose generalized parity. From this expression, we can find an upper bound on the individual entries in the Yukawa matrix, of the form
\begin{equation}
	| (\tilde y_d)_{i j} | \lesssim \left( \frac{2 m_{d_i} m_{D_j}}{v v'} \right)^{1/2} \sim  \left( \frac{m_{d_i} M}{v v'} \right)^{1/2} \ .
\label{eq:ymax}
\end{equation}

As advertised in section~\ref{sec:fermionmasses}, bringing the full $6 \times 6$ matrix of Eq.(\ref{eq:M6x6}) into diagonal form requires a further transformation that mixes the $SU(2)$-doublet and singlet fields, as specified in Eq.(\ref{eq:fermionrot1}). In terms of the $\tilde y_d$ and $\tilde y'_d$ couplings defined earlier, the $3 \times 3$ blocks appearing in Eq.(\ref{eq:fermionrot1}) can be written as
\begin{equation}
	\epsilon_d = \frac{v}{\sqrt{2}} \mathbb{m}_D^{-1} \tilde y_d^T , \qquad \qquad {\rm and} \qquad \qquad \epsilon'_d = \frac{v'}{\sqrt{2}} \mathbb{m}_D^{-1} \tilde y'^\dagger_d ,
\label{eq:epsilond}
\end{equation}
whose entries are of $\co (v / M)$ and $\co(v' / M)$ respectively. When generalized parity is only broken by the different vev's in the SM and mirror sectors, we have $\epsilon'_d = (v' / v) \epsilon_d^*$.

\subsubsection*{Up-type quarks} 

The diagonalization procedure in the up-quark sector is analogous to that for down-type quarks and leptons, although this time accommodating for the singularities of the third generation for which the see-saw mechanism cannot be implemented.

As before, at zeroth order in $v^{(\prime)}/M$, we perform transformations of the form 
\begin{equation}
	u \rightarrow \co^\dagger_u u , \ u' \rightarrow \co^\dagger_{u'} u' , \qquad {\rm and} \qquad U \rightarrow \co^\dagger_U U , \ U' \rightarrow \co^\dagger_{U'} U' .
\end{equation}
On the one hand, the matrices $\co_U$ and $\co_{U'}$ must be chosen such that the vector-like mass matrix $\cm_u$ is brought into diagonal form. In this case, we make the assumption that two of the eigenvalues of $\cm_u$ are $m_{U_1}, m_{U_2} \sim M$, whereas the third one is much smaller, and for simplicity we will take it to vanish in what follows. On the other hand, the matrices $\co_u$ and $\co_{u'}$ must now be such that
\begin{equation}
	(\tilde y'^*_u \hat{\mathbb{m}}_U^{-1} \tilde y^T_u)_{ij} = \delta_{ij} (\tilde y'^*_u \hat{\mathbb{m}}_U^{-1} \tilde y^T_u)_{ii} , \qquad \quad {\rm and} \quad \qquad
	(\tilde y_u)_{i3} = (\tilde y'_u)_{i3} = 0
\end{equation}
for $i, j = 1, 2$, and where $\hat{\mathbb{m}}_U^{-1} \equiv {\rm diag} (m_{U_1}^{-1}, m_{U_2}^{-1}, 0)$, and the $\tilde y_u$ and $\tilde y'_u$ matrices are defined as in Eq.(\ref{eq:deftildey}).
Moreover, we define $y_t \equiv \tilde y_{33}$, and $y_{t'} \equiv \tilde y'^*_{33}$, which we may choose to be real and positive.

With this preliminaries, the tree-level mass eigenvalues in the top sector read
\begin{equation}
	m_t = \frac{y_t}{\sqrt{2}} v \ , \qquad {\rm and} \qquad m_{t'} = \frac{y_{t'}}{\sqrt{2}} v'  ,
\end{equation}
with $y_{t'} = y_t$ if we impose generalized parity. For the first and second generation, we have instead
\begin{equation}
	m_{u_i} = \frac{v v'}{2} \sum_{j=1}^2 \frac{(\tilde y'^*_u)_{ij} (\tilde y_u)_{ij}}{m_{U_j}} = \frac{v v'}{2} \sum_{j=1}^2 \frac{ | (\tilde y_u)_{ij} |^2 }{m_{U_j}} ,
\end{equation}
where the last step holds provided we impose generalized parity. As before, we can now obtain an upper bound on the individual Yukawa entries, of the form
\begin{equation}
	| (\tilde y_u)_{i j} | \lesssim \left( \frac{2 m_{u_i} m_{U_j}}{v v'} \right)^{1/2} \sim  \left( \frac{m_{u_i} M}{v v'} \right)^{1/2} \qquad {\rm for} \qquad i,j=1,2 .
\label{eq:yumax}
\end{equation}

A further transformation mixing the $SU(2)$-singlet and doublet components is again necessary in order to diagonalize the full $6 \times 6$ mass matrix, which can be written as in Eq.(\ref{eq:fermionrot1}). The corresponding $\epsilon_u$ and $\epsilon'_u$ blocks are now given by
\begin{equation} \begin{aligned}
	(\epsilon_u)_{ij} & = \frac{v}{\sqrt{2}} ( \hat{\mathbb{m}}_U^{-1} \tilde y^T )_{ij} \ , \qquad
	& (\epsilon_u)_{3j} & = - \frac{v v'}{2} \frac{m_{t'} (\tilde y'^* \hat{\mathbb{m}}_U^{-1} \tilde y^T)_{3j}}{m_{t'}^2 - \delta_{j3} m_t^2} \ , \\
	(\epsilon'_u)_{ij} & = \frac{v'}{\sqrt{2}} ( \hat{\mathbb{m}}_U^{-1} \tilde y'^T )^*_{ij} \ , \qquad
	& (\epsilon'_u)_{3j} & = - \frac{v v'}{2} \frac{m_t (\tilde y^* \hat{\mathbb{m}}_U^{-1} \tilde y'^T)^*_{3j}}{m_t^2 - \delta_{j3} m_{t'}^2} \ .
\label{eq:epsilonu}
\end{aligned} \end{equation}
for $i=1,2$ and $j=1,2,3$. Just as in the down-quark sector, if generalized parity is only broken by the difference between $v$ and $v'$, we have $\epsilon'_u = (v' / v) \epsilon^*_u$.

\section{Radiatively induced EDM}

\subsection{One-loop EDM}
\label{sec:appEDM}

We will now present a calculation of the one-loop correction to the EDM of elementary charged fermions that arises under the assumption that parity is only broken softly, both in the scalar potential and through the presence of non-hermitian vector-like masses for the $SU(2)$-singlets. The relevant diagrams are those featured in figure~\ref{fig:1loopEDM}. We will concentrate first on diagrams where either $h$ or $h'$ propagate inside the loop.

In full generality, a Dirac fermion $f$ that interacts with another fermion $\psi$, and a neutral scalar $\phi$ through Yukawa couplings of the form
\begin{equation}
	\cl \supset L (\bar f_R \psi_L) \phi + R (\bar f_L \psi_R) \phi + {\rm h.c.},
\label{eq:masterL}
\end{equation}
will receive a one-loop EDM given by
\begin{equation}
	\frac{d_f}{e} = \frac{Q_\psi}{16 \pi^2} \frac{m_\psi}{m^2_\phi} A(r) \Imag \left( L R^* \right),
\end{equation}
where $r = m_\psi^2 / m_\phi^2$, and the loop function $A$ is given by
\begin{equation}
	A(r) = \frac{1}{2(1-r)^2} \left( 3 - r + \frac{2 \log r}{1-r} \right).
\end{equation}
In the model we are considering, the Yukawa interactions involving both light and heavy fermions can be written as
\begin{equation}
	\cl \supset - \sum_{s=h, h'} s \left(		\bar d_R \hat \eta^s d_L		+ \bar d_R \hat \beta^s D_L	+ \bar D_R \hat \gamma^s d_L	+ \bar D_R \hat \delta^s D_L	\right)	+ \rm {h.c.},
\label{eq:yukawasd}
\end{equation}
where we are using notation specific to the down-quark sector, but analogous expressions apply for up-quarks and leptons (although the specific form of the Yukawa matrices will differ).
It will be convenient to write the above matrices as $\hat \omega^s = R_{1s} \omega^h + R_{2s} \omega^{h'}$ (for $\omega = \eta, \beta, \gamma, \delta$), where $R_{11}=R_{22} = \cos \alpha$ and $R_{12} = - R_{21} = \sin \alpha$, with $\alpha \sim v/v'$ the mixing angle the Higgs sector. For the down-quark sector, the $\omega$ matrices can be conveniently written as follows
\begin{equation}
	\eta^h = - \frac{\mathbb{m}_d}{v},	\qquad	\beta^h = - \frac{1}{\sqrt{2}} \left( \mathbb{m}_d \tilde y^* \mathbb{m}_D^{-1} \right),	\qquad
	\gamma^h = \frac{\tilde y^T}{\sqrt{2}},					\qquad	\delta^h = \frac{v}{2} \left( \tilde y^T \tilde y^* \mathbb{m}_D^{-1} \right),
\label{eq:omegas_h}
\end{equation}
and
\begin{equation}
	\eta^{h'} = - \frac{\mathbb{m}_d}{v'},	 				\qquad	\beta^{h'} = \frac{\tilde y'^*}{\sqrt{2}},		\qquad
	\gamma^{h'} = - \frac{1}{\sqrt{2}} \mathbb{m}_D^{-1} \tilde y'^T \mathbb{m}_d,		\qquad	\delta^{h'} = \frac{v'}{2} \left( \mathbb{m}_D^{-1} \tilde y'^T \tilde y'^* \right) .
\label{eq:omegas_hp}
\end{equation}

The one-loop correction to the EDM of one of the SM-like quarks, $d_i$, is dominated by diagrams where the heavy mirror quarks propagate inside the loop. Since $m_{D_j} \sim M \gg m_h, m_{h'}$, we can expand the loop function $A(r)$ in the limit $r \gg 1$. Keeping the first two terms, we find
\begin{equation}
	\frac{d_{d_i}}{e} \simeq - \sum_{j,s} \frac{Q_d}{32 \pi^2 m_{D_j}} \left( 1 - \frac{m_s^2}{m_{D_{j}}^2} \right) \Imag \left( \hat \beta^s_{ij} \hat \gamma^s_{ji} \right) .
\label{eq:ddiv1}
\end{equation}
The last factor in the previous expression can be written as
\begin{equation}
	\Imag( \hat \beta^s_{ij} \hat \gamma^s_{ji} ) = R_{1s} R_{2s} \Imag ( \beta^h_{ij} \gamma^{h'}_{ji} + \beta^{h'}_{ij} \gamma^h_{ji} ),
\end{equation}
where we have taken into account that $\beta^h_{ij} \gamma^h_{ji}, \beta^{h'}_{ij} \gamma^{h'}_{ji} \in \mathbb{R}$, so those combinations don't appear on the right-hand-side. When summing over $s$ in Eq.(\ref{eq:ddiv1}), the contribution from the leading term in the $m_s^2 / m_{D_j}^2 \ll 1$ expansion vanishes since $\sum_s R_{1s} R_{2s} = 0$.
The leading contribution to $d_{d_i}$ then reads
\begin{equation} \begin{split}
	\frac{d_{d_i}}{e} 	& \simeq \sum_{j,s} \frac{Q_d}{32 \pi^2} \frac{m_s^2}{m_{D_{j}}^3} R_{1s} R_{2s} \Imag \left( \beta^h_{ij} \gamma^{h'}_{ji} + \beta^{h'}_{ij} \gamma^h_{ji} \right) \\
					& \simeq \sum_{j} \frac{Q_d}{32 \pi^2} \frac{m_{h'}^2}{m_{D_{j}}^3} \sin \alpha \Imag \left( \beta^{h'}_{ij} \gamma^h_{ji} \right) ,
\label{eq:ddiv2}
\end{split} \end{equation}
where in the last step we have neglected the first term in parenthesis since it is suppressed by a factor of $\co( (v v'/M^2)^2 )$ with respect to the second, and we have only kept the contribution from $h'$, since the contribution from $h$ is suppressed by an additional factor of $m_h^2 / m_{h'}^2$.

From Eq.(\ref{eq:omegas_h}) and (\ref{eq:omegas_hp}), we find
\begin{equation}
	\beta^{h'}_{ij} \gamma^h_{ji} = \frac{1}{2} \tilde y'^*_{ij} \tilde y_{ij} .
\label{eq:ypy}
\end{equation}
When generalized parity is a good symmetry, we have $\tilde y' = \tilde y$, and therefore the above term is real, in turn leading to a vanishing EDM. In the presence of soft breaking through non-hermitian vector-like masses, the relationship $\tilde y' = \tilde y$ no longer holds, even if the Yukawa couplings in the flavor basis remain identical since the breaking is soft. To see that the equality of the Yukawa couplings in the SM and mirror sectors no longer holds in the mass basis, it is useful to remind ourselves of the fermion mass diagonalization procedure discussed in appendix \ref{sec:appfermions}. When generalized parity remains unbroken, the unitary matrices of Eq.(\ref{eq:rotationsd}) are such that $\co_F = \co_{F'}^*$ and $\co_{f'} = \co_f^*$. However, the non-hermiticity of the vector-like mass matrix means the unitary matrices needed to bring the $3 \times 3$ matrices of Eq.(\ref{eq:dquarksmasses}) into real diagonal form will no longer satisfy this simple relation.
Instead, writing the new vector-like mass matrices as $\cm_f + i \Delta \cm_f$, with both $\cm_f$ and $\Delta \cm_f$ hermitian, the new unitary matrices  are modified as follows
\begin{equation} \begin{aligned}
	\co_{F'} & \rightarrow \tilde \co_{F'} = \co_{F'} + \Delta_{F'} ,	\qquad \qquad		& & \co_F \rightarrow \tilde \co_F = \co^*_{F'} - \Delta^*_{F'}, \\
	\co_f & \rightarrow \tilde \co_f = \co_f + \Delta_f ,			\qquad \qquad		& & \co_{f'} \rightarrow \tilde \co_{f'} = \co^*_f - \Delta^*_f .
\end{aligned} \end{equation}
The $\Delta$ matrices arise at $\co \left( |\Delta \cm| / M \right)$, and are given by
\begin{equation}
	\left( \Delta_{F'} \right)_{ij} = i \sum_{k \neq i} \frac{[\Delta \tilde \cm, \mathbb{m}_F ]_{ik}}{m^2_{F_i} - m^2_{F_k}} \left( \co_{F'} \right)_{kj} , \quad {\rm and} \quad
	\left( \Delta_f \right)_{ij} = i \sum_{k \neq i} \frac{[ \Delta \tilde m, \mathbb{m}_f ]_{ik}}{m^2_{f_i} - m^2_{f_k}} \left( \co_f \right)_{kj} ,
\end{equation}
where
\begin{equation}
	\Delta \tilde \cm \equiv \co_{F'} \Delta \cm \co_{F'}^\dagger	\qquad {\rm and} \qquad
	\Delta \tilde m \equiv \frac{v v'}{2} \tilde y^* \mathbb{m}_F^{-1} \Delta \tilde \cm \mathbb{m}_F^{-1} \tilde y^T .
\end{equation}

Using the above expressions in the definition of $\tilde y$, it is possible to write
\begin{equation}
	\tilde y'^* = \tilde y^* (\mathbb{1} + \xi) ,
\end{equation}
where $\xi$ is a matrix with entries of $\co (|\Delta \cm| / M)$. Explicitly, after some massaging,
\begin{equation}
	\tilde y'^*_{ij} = \sum_l \tilde y^*_{il} \left( \delta_{lj} + i \frac{\Delta \tilde \cm_{lj}}{m_{F_l} + m_{F_j}} (1 - \delta_{lj}) + i \frac{v v'}{2} \sum_{n, k \neq i} \frac{\tilde y^*_{kj} \tilde y_{kn}}{m_{f_i} + m_{f_k}} \frac{\Delta \tilde \cm_{ln}}{m_{F_l} m_{F_n}} \right) .
\end{equation}
In total:
\begin{equation}
	\Imag (\tilde y'^*_{ij} \tilde y_{ij}) = |\tilde y_{ij}|^2 \times \co \left( \frac{\Delta \cm}{M} \right) .
\end{equation}
Plugging this back into Eq.(\ref{eq:ddiv2}), we have
\begin{equation}
	\frac{d_{d_i}}{e} \simeq \sum_j \frac{Q_d}{32 \pi^2} \frac{m_{h'}^2}{m_{D_j}^3} \sin \alpha \frac{1}{2} \Imag (\tilde y'^*_{ij} \tilde y_{ij})
				\simeq \frac{n_d Q_d}{32 \pi^2} \frac{m_{d_i}}{M^2} \times \co \left( \frac{\Delta \cm}{M} \right) ,
\label{eq:ddifinal}
\end{equation}
with $n_d = 3$ the number of mirror fermions appearing at the see-saw scale in the down-quark sector. The above expression also applies to the lepton sector, after making the obvious substitutions. In the up-quark sector, the expressions for the Yukawa couplings are somewhat different to those in Eq.(\ref{eq:omegas_h}) and (\ref{eq:omegas_hp}), but can be similarly found by following the flavor-to-mass-basis rotation procedure outlined in section \ref{sec:appfermions}. In the end, diagrams where $h'$ and the mirror partners of the $u$ and $c$ quarks propagate inside the loop give the leading contribution to the one-loop EDM. Thus, the above expression also applies for the up-quark sector, this time with $n_u = 2$ instead.

Additional contributions arise from diagrams where $Z$ and $Z'$ propagate inside the loop (see figure \ref{fig:1loopEDM}). In this case, the leading contribution arises from diagrams involving $Z'$ as well as heavy mirror fermions. In total, the final result is parametrically the same as that in Eq.(\ref{eq:ddifinal}), except for an additional suppression by a factor of $g^2 \sin^2 \theta_w$.

Although the potential one-loop correction to $\bar \theta$ that could arise as a result of the soft breaking through non-hermitian vector-like masses was already shown to vanish in \cite{Babu:1989rb}, this can also be seen from the calculation we have just performed. The relevant diagrams contributing to the quark mass matrix, and therefore to $\bar \theta$, are those of figure \ref{fig:1loopEDM}, minus the external photon line. So although the appropriate loop function will be different, the overall correction will be similarly proportional to Eq.(\ref{eq:ypy}). Using Eq.(\ref{eq:deffmd}) to rewrite $\tilde y'$ in terms of $\tilde y$, and the diagonal mass matrices, we find
\begin{equation}
	\Imag(\delta m_{d_i}) \propto \Imag (\tilde y'^*_{ij} \tilde y_{ij}) \propto m_{d_i} \Imag ( (\tilde y^{-1})_{ji} \tilde y_{ij}) .
\end{equation}
As a result, the corresponding contribution to $\bar \theta$ from the down-quark sector reads
\begin{equation}
	\sum_i \frac{\Imag(\delta m_{d_i})}{m_{d_i}} \propto \Imag \left(  \sum_i (\tilde y^{-1})_{ji} \tilde y_{ij} \right) = 0 .
\end{equation}
Notice the sum over quark flavors is crucial in the above cancellation.

\subsection{One-loop $\bar \theta$}
\label{sec:apptheta}

The calculation of the one-loop correction to the quark mass matrix, and, in turn, to $\bar \theta$, proceeds along similar lines to the EDM calculation we have just discussed. The leading contribution to $\bar \theta$ comes from corrections to the light quark masses, and it is due to diagrams where either $h'$ or $\phi$ propagate inside the loop.

Rotating from the gauge to the mass basis in the scalar sector requires performing a transformation $s_i \rightarrow R_{ij} s_j$, with $s_i = \{h, h', \phi\}$, and $R$ is a $3 \times 3$ orthogonal matrix that we parametrize in terms of the various mixing angles as
\begin{equation}
	R = \begin{pmatrix}
	{\rm c}_\alpha {\rm c}_\beta 	& & & {\rm c}_\alpha  {\rm s}_\beta {\rm s}_\gamma + {\rm s}_\alpha {\rm c}_\gamma 	& & & - {\rm c}_\alpha  {\rm s}_\beta {\rm c}_\gamma + {\rm s}_\alpha {\rm s}_\gamma\\ 
	- {\rm s}_\alpha {\rm c}_\beta 	& & & - {\rm s}_\alpha  {\rm s}_\beta {\rm s}_\gamma + {\rm c}_\alpha {\rm c}_\gamma 	& & & {\rm s}_\alpha  {\rm s}_\beta {\rm c}_\gamma + {\rm c}_\alpha {\rm s}_\gamma\\ 
	{\rm s}_\beta 				& & & -  {\rm c}_\beta {\rm s}_\gamma									 	& & & {\rm c}_\beta {\rm c}_\gamma
	\end{pmatrix}
\end{equation}
where ${\rm c}_\alpha = \cos \alpha$, ${\rm s}_\alpha = \sin \alpha$, etc. Parametrically, we expect ${\rm c}_\alpha \sim {\rm c}_\beta \sim {\rm c}_\gamma = \co(1)$, whereas ${\rm s}_\alpha \sim v/v'$, ${\rm s}_\beta \sim v/v_\phi$, and ${\rm s}_\gamma \sim v_\phi/v'$.
In the down-quark sector, the Yukawa interactions of Eq.(\ref{eq:yukawasd}) need to be extended to include $\phi$ in the sum, and the $\hat \omega^s$ matrices are now given by
\begin{equation}
	\hat \omega^s = R_{1s} \omega^h + R_{2s} \omega^{h'} + R_{3s} \omega^\phi	\qquad {\rm for} \qquad	\omega = \eta, \beta, \gamma, \delta.
\end{equation}
The expressions for $\omega^h$ and $\omega^{h'}$ are as in Eq.(\ref{eq:omegas_h}) and (\ref{eq:omegas_hp}), whereas for $\phi$ we have
\begin{equation} \begin{aligned}
	\eta^\phi & = - i \frac{v v'}{2} \left( \tilde y'^* \mathbb{m}_D^{-1} \tilde {\bar y} \mathbb{m}_D^{-1} \tilde y^T \right), \qquad
	&\beta^\phi & = \frac{i v'}{\sqrt{2}} \left( \tilde y'^* \mathbb{m}_D^{-1} \tilde {\bar y} \right), \\
	\gamma^\phi & = \frac{i v}{\sqrt{2}} \left( \tilde {\bar y} \mathbb{m}_D^{-1} \tilde y^T \right), \qquad
	&\delta_\phi & = - i \tilde {\bar y} ,
\end{aligned} \end{equation}
where $\tilde {\bar y} \equiv \co_{F'} \bar y \co_{F'}^\dagger$, as usual.

In the notation of Eq.(\ref{eq:masterL}), the one-loop correction to $\Imag (\delta m_f)$ is given by
\begin{equation}
	\Imag (\delta m_f) = \frac{m_\psi}{16 \pi^2} \mathcal{F} (m_\psi, m_\phi) \Imag(L R^*) ,
\end{equation}
where the loop function $ \mathcal{F}$ now reads
\begin{equation}
	\mathcal{F} (m_\psi, m_\phi) = \frac{1}{m_\psi^2 - m_\phi^2} \left[ m_\psi^2 \left( \log \frac{m_\psi^2}{\mu^2} - 1 \right) - m_\phi^2 \left( \log \frac{m_\phi^2}{\mu^2} - 1 \right) \right] .
\end{equation}
For the case at hand, the leading one-loop correction to the mass of the SM-like fermions involves diagrams where the heavy mirror partners appearing at scale $M$ propagate inside the loop. Specifically, in the down-quark sector, we have
\begin{equation}
	\Imag (\delta m_{d_i}) = \sum_{s, j} \frac{m_{D_j}}{16 \pi^2} \mathcal{F} (m_{D_j}, m_s) \Imag (\hat \beta^s_{ij} \hat \gamma^s_{ji}) .
\end{equation}
In analogy to the discussion in the previous section, the leading term in $\mathcal{F}$ in the limit $m_{D_j} \gg m_s$ is independent of $m_s$, and its contribution to $\Imag (\delta m_f)$ vanishes as a result of the orthogonality of the mixing matrix in the scalar sector.
The leading correction to $\Imag (\delta m_{d_i})$ then reads
\begin{equation} \begin{split}
	\Imag (\delta m_{d_i}) 	&\simeq \sum_{j, s} \frac{m_s^2}{16 \pi^2 m_{D_j}} \log\frac{m_{D_j}^2}{m_s^2} \Imag (\hat \beta^s_{ij} \hat \gamma^s_{ji}) \\
						&\simeq \sum_{j, s=h', \phi} \frac{m_s^2}{16 \pi^2 m_{D_j}} \log\frac{m_{D_j}^2}{m_s^2}
							\Imag ({\rm s}_\gamma \beta^{h'}_{ij} \gamma^\phi_{ji} + {\rm s}_\gamma {\rm s}_\alpha \beta^\phi_{ij} \gamma^h_{ji}) ,
\end{split} \end{equation}
where in the last step we have neglected the contribution from $h$, which is suppressed by a factor of $m_h^2 / m_s^2$ compared to that from $h'$ and $\phi$.
The two terms inside the parenthesis are given by
\begin{equation}
	{\rm s}_\gamma \Imag ( \beta^{h'}_{ij} \gamma^\phi_{ji} )
	= {\rm s}_\gamma \frac{v}{2} \frac{\tilde y'^*_{ij} \tilde {\bar y}_{jk} \tilde y_{ik}}{m_{D_k}}
	\sim {\rm s}_\gamma \frac{m_{d_i} \bar y }{v'} \sim \frac{ v_\phi m_{d_i} \bar y }{v'^2} ,
\end{equation}
where in the last step we have substituted ${\rm s}_\gamma \sim v_\phi / v'$, as we expect when $v_\phi \lesssim v'$, and
\begin{equation}
	{\rm s}_\gamma {\rm s}_\alpha \Imag ( \beta^\phi_{ij} \gamma^h_{ji} )
	= {\rm s}_\gamma {\rm s}_\alpha \frac{v'}{2} \frac{\tilde y'^*_{ik} \tilde {\bar y}_{kj} \tilde y_{ij}}{m_{D_k}}
	\sim {\rm s}_\gamma {\rm s}_\alpha \frac{m_{d_i} \bar y }{v} \sim \frac{ v_\phi m_{d_i} \bar y }{v'^2} .
\end{equation}
Both terms are therefore of the same order. When $v_\phi \lesssim v'$, the contribution from $\phi$ to $\Imag (\delta m_{d_i}) $ is subleading to that from $h'$.
Setting $m_{h'} \simeq \sqrt{2 \lambda} v' \sim v'$, we then have
\begin{equation}
	\Imag (\delta m_{d_i}) \sim \frac{m_{d_i}}{16 \pi^2} \frac{\bar y v_\phi}{M} \log \frac{M^2}{m_{h'}^2},
\end{equation}
and the contribution to $\bar \theta$ from the down-quark sector reads
\begin{equation}
	\bar \theta \simeq \sum_i \frac{\Imag (\delta m_{d_i})}{m_{d_i}} \sim \frac{1}{16 \pi^2} \frac{\bar y v_\phi}{M} \log \frac{M^2}{m_{h'}^2} .
\end{equation}
The above expression agrees with the parametric estimate presented in section \ref{sec:PandCP}, except for the log factor that is not captured in our spurion analysis.

\section{Kaon mixing}
\label{app:flavor}

The $\Delta m_K$ and $|\epsilon_K|$ parameters characterizing the kaon sector can be written as
\begin{equation}
	\Delta m_K = 2 {\rm Re} (m^K_{12}), \qquad {\rm and} \qquad |\epsilon_K| = \frac{\kappa_\epsilon |\Imag (m^K_{12})|}{\sqrt{2} \Delta m_K} ,
\end{equation} 
where $m^K_{12} \equiv \frac{1}{2 m_K} \langle K^0 | \mathcal{H}_{\rm eff} | \bar K^0 \rangle$, and $\mathcal{H}_{\rm eff}$ refers to the effective hamiltonian appropriate to describe kaon mixing.

In the SM, $\mathcal{H}_{\rm eff}$ is generated at one-loop through box diagrams involving two $W$ gauge bosons. The corresponding contribution reads
\begin{equation}
	\mathcal{H}_{\rm eff} \supset - \frac{G_F^2 m_W^2}{4 \pi^2} (\bar d_L \gamma_\mu s_L) (\bar d_L \gamma^\mu s_L) \sum_{\alpha, \beta} \lambda_\alpha \lambda_\beta F(x_\alpha, x_\beta) + {\rm h.c.},
\label{eq:HeffLL}
\end{equation}
where the loop function $F$ is given by \cite{Inami:1980fz}
\begin{equation} \begin{aligned}
	F (x_\alpha, x_\beta)		& = \frac{x_\alpha^2 \log x_\alpha}{(x_\beta - x_\alpha) (1-x_\alpha)^2} \left( 1 - 2 x_\beta + \frac{x_\alpha x_\beta}{4} \right) + \{ x_\alpha \leftrightarrow x_\beta \} \\
						& + \frac{1}{(1-x_\alpha)(1-x_\beta)} \left( \frac{7 x_\alpha x_\beta}{4} - 1 \right) ,
\end{aligned} \end{equation}
and $\lambda_\alpha = V_{\alpha d}^* V_{\alpha s}$ for $\alpha = u, c, t$. In the present model, the sum over $\alpha$ and $\beta$ in Eq.(\ref{eq:HeffLL}) must be extended to include the additional members of the up-quark sector. The corresponding couplings can be read off from Eq.(\ref{eq:LWups}), and are given by
\begin{equation}
	\lambda_\alpha = \Delta V_{\alpha d}^* \Delta V_{\alpha s} \qquad {\rm for} \qquad \alpha = U, C, T .
\end{equation}

An additional contribution to $\mathcal{H}_{\rm eff}$ arises from diagrams involving one $W$ and one $W'$. In this case:
\begin{equation}
	\mathcal{H}_{\rm eff} \supset - \frac{G_F^2 m_W^2}{4 \pi^2} \beta (\bar d_R s_L) (\bar d_L s_R) \sum_{\alpha, \beta} \lambda^{LR}_\alpha \lambda^{RL}_\beta \tilde F (\beta, x_\alpha, x_\beta) + {\rm h.c.},
\end{equation}
where $\beta \equiv m_W^2 / m_{W'}^2 = v^2 / v'^2$, and the loop function now reads \cite{Ecker:1985vv}
\begin{equation}
	\tilde F (\beta, x_\alpha, x_\beta) = \sqrt{x_\alpha x_\beta} \left\{ (1 + \beta) I_2 (x_\alpha, x_\beta, \beta) - (4 + \beta x_\alpha x_\beta) I_1 (x_\alpha, x_\beta, \beta) \right\},
\end{equation}
with
\begin{equation} \begin{aligned}
	I_1 (\beta, x_\alpha, x_\beta) & = \frac{x_\alpha \log x_\alpha}{(1- x_\alpha) (1 - \beta x_\alpha) (x_\alpha - x_\beta)}
		+ \{ x_\alpha \leftrightarrow x_\beta \}
		- \frac{\beta \log \beta}{(1 - \beta) (1 - \beta x_\alpha) (1 - \beta x_\beta)} ,\\
	I_2 (\beta, x_\alpha, x_\beta) & = \frac{x^2_\alpha \log x_\alpha}{(1- x_\alpha) (1 - \beta x_\alpha) (x_\alpha - x_\beta)}
		+ \{ x_\alpha \leftrightarrow x_\beta \}
		- \frac{\log \beta}{(1 - \beta) (1 - \beta x_\alpha) (1 - \beta x_\beta)} .
\end{aligned} \end{equation}
The relevant couplings follow from the interactions in Eq.(\ref{eq:LWups}) and (\ref{eq:LWpups}). For the $u$ and $c$ quarks, we have $\lambda^{LR}_\alpha = \lambda^{RL}_\alpha = \lambda_\alpha$, whereas for their heavy partners
\begin{equation}
	\lambda^{LR}_\alpha = \Delta V^*_{\alpha d} \Delta V'_{\alpha s},
	\qquad {\rm and} \qquad \lambda^{RL}_\alpha = \Delta V'^*_{\alpha d} \Delta V_{\alpha s},
	\qquad {\rm for} \qquad \alpha=U,C.
\end{equation}
In the top sector, on the other hand, we have
\begin{equation}
	\lambda^{LR}_t = V^*_{td} \Delta V'_{3 s}, \ \lambda^{RL}_t = \Delta V'^*_{3d} V_{t s},
	\qquad {\rm and} \qquad
	\lambda^{LR}_T = \Delta V^*_{3d} V_{t s}, \ \lambda^{RL}_T = V^*_{td} \Delta V_{3 s}.
\end{equation}


\bibliography{strongcp_refs}
	
\end{document}